\documentclass[useAMS]{mn2e}
\usepackage{graphicx}
\usepackage{epsfig}
\usepackage{amssymb}
\usepackage{threeparttable}
\usepackage{multicol}
\usepackage{multirow}

\def\ltsima{$\; \buildrel < \over \sim \;$}
\def\lsim{\lower.5ex\hbox{\ltsima}}
\def\gtsima{$\; \buildrel > \over \sim \;$}
\def\gsim{\lower.5ex\hbox{\gtsima}}

\newcommand{\be}{\begin{equation}}
\newcommand{\en}{\end{equation}}

\def\nh{\hbox{$N_{\rm H}$}}

\def\ergscm2   {erg  \, s$^{-1}$ \, cm$^{-2}$}

\def\aa {1E\,1547$-$5408}

\def\wes{CXOU\,J1647$-$4552}

\def\sgre{SGR\,0501$+$4516}
\def\lowba{SGR\,0418$+$5729}

\def\3xmm{3XMM\,J1852$+$0033}

\def\galcen{SGR\,1745$-$2900}

\newcommand{\AXAF}{{\em Chandra}}

\newcommand{\XMM}{{\em XMM--Newton}}

\newcommand{\swift}{{\em Swift}}

\newcommand{\INT}{{\em INTEGRAL}}

\begin{document}

\title[The 1.5-year X-ray outburst of SGR\,J1745-2900]{The X-ray outburst of the Galactic Centre magnetar SGR\,J1745-2900 during the first 1.5 year}
\author[F. Coti Zelati et al.]
{F. Coti Zelati,$^{1,2,3}$\thanks{E-mail: francesco.cotizelati@brera.inaf.it} N. Rea,$^{2,4}$ A. Papitto,$^{4}$ D. Vigan\`o,$^{4}$ J. A. Pons,$^{5}$ \newauthor R. Turolla,$^{6,7}$ P. Esposito,$^{8,9}$ D. Haggard,$^{10}$ F. K. Baganoff,$^{11}$ G. Ponti,$^{12}$ \newauthor G. L. Israel,$^{13}$ S. Campana,$^{3}$ D. F. Torres,$^{4,14}$ A. Tiengo,$^{8,15,16}$ S. Mereghetti,$^{8}$ \newauthor R. Perna,$^{17}$ S. Zane,$^{7}$  R. P. Mignani,$^{8,18}$ A. Possenti,$^{19}$ L. Stella$^{13}$\\
$^{1}$ Universit\`a dell'Insubria, via Valleggio 11, I-22100 Como, Italy\\
$^{2}$ Anton Pannekoek Institute for Astronomy, University of Amsterdam, Postbus 94249,  NL-1090-GE Amsterdam, The Netherlands\\
$^{3}$ INAF -- Osservatorio Astronomico di Brera, via Bianchi 46, I-23807 Merate (LC), Italy\\
$^{4}$ Institute of Space Sciences (ICE, CSIC--IEEC), Carrer de Can Magrans, S/N, 08193, Barcelona, Spain\\
$^{5}$ Departament de Fisica Aplicada, Universitat d'Alacant, Ap. Correus 99, E-03080 Alacant, Spain\\
$^{6}$ Dipartimento di Fisica e Astronomia, Universit\`a di Padova, via F. Marzolo 8, I-35131 Padova, Italy\\
$^{7}$ Mullard Space Science Laboratory, University College London, Holmbury St. Mary, Dorking, Surrey RH5 6NT, UK\\
$^{8}$ INAF -- Istituto di Astrofisica Spaziale e Fisica Cosmica, via E. Bassini 15, I-20133 Milano, Italy\\
$^{9}$ Harvard--Smithsonian Center for Astrophysics, 60 Garden Street, Cambridge, MA 02138, USA\\
$^{10}$ Department of Physics and Astronomy, Amherst College, Amherst, MA 01002-5000, USA\\ 
$^{11}$ Kavli Institute for Astrophysics and Space Research, Massachusetts Institute of Technology, Cambridge, MA 02139, USA\\
$^{12}$ Max Planck Institute fur Extraterrestriche Physik, Giessenbachstrasse, D-85748 Garching, Germany\\
$^{13}$ INAF -- Osservatorio Astronomico di Roma, via Frascati 33, I-00040 Monteporzio Catone, Roma, Italy\\
$^{14}$ Instituci\'o Catalana de Recerca i Estudis Avan\c{c}ats (ICREA) Barcelona, Spain\\
$^{15}$ Istituto Universitario di Studi Superiori, piazza della Vittoria 15, I-27100 Pavia, Italy\\
$^{16}$ Istituto Nazionale di Fisica Nucleare, Sezione di Pavia, via A. Bassi 6, I-27100 Pavia, Italy\\
$^{17}$ Department of Physics and Astronomy, Stony Brook University, Stony Brook, NY 11794, USA\\
$^{18}$ Kepler Institute of Astronomy, University of Zielona G�ora, Lubuska 2, 65-265, Zielona G\'ora, Poland\\
$^{19}$ INAF -- Osservatorio Astronomico di Cagliari, via della Scienza 5, 09047, Selargius, Cagliari, Italy
}

\date{}
\maketitle

\begin{abstract}
In 2013 April a new magnetar, \galcen , was discovered as it entered 
an outburst, at only 2.4 arcsec angular distance from the supermassive black hole at the 
Centre of the Milky Way, Sagittarius A$^*$. \galcen\ has a surface dipolar magnetic 
field of $\sim 2\times10^{14}$\,G, and it is the neutron star closest to a black 
hole ever observed. The new source was detected both in the radio and 
X-ray bands, with a peak X-ray luminosity $L_X \sim  5\times 10^{35}$\,erg~s$^{-1}$.
Here we report on the long-term {\AXAF} (25 observations) and {\XMM} (8 observations) 
X-ray monitoring campaign of \galcen\, from the onset of the outburst in April 2013 
until September 2014. This unprecedented dataset allows us to refine the timing
properties of the source, as well as to study the outburst spectral
evolution as a function of time and rotational phase. Our timing analysis confirms the 
increase in the spin period derivative by a factor of $\sim$2 around June 2013, and 
reveals that a further increase occurred between 2013 Oct 30 and 2014 Feb 21. We 
find that the period derivative changed from 6.6$\times10^{-12}$ s~s$^{-1}$ to 
3.3$\times10^{-11}$ s~s$^{-1}$ in 1.5 yr. On the other hand, this magnetar shows a 
slow flux decay compared to other magnetars and a rather inefficient surface cooling. 
In particular, starquake-induced crustal cooling models alone have difficulty in 
explaining the high luminosity of the source for the first $\sim$200\,days of its outburst, 
and additional heating of the star surface from currents flowing in a twisted magnetic 
bundle is probably playing an important role in the outburst evolution.
\end{abstract}

\begin{keywords}
Galaxy: centre -- stars: magnetars -- X-rays: individual: SGR\,J1745-2900.
\end{keywords}

\section{Introduction}

%%%%%%%%%%%%%%%%%%%%%%%%%%%%%%%%%%%%%%%%%%%%%%%%%%%%%%%%%%%%%%%%%%%%%%%%%%%%%%%%%%%%%%%%%%
\begin{table*}
\caption{Log of {\AXAF}/ACIS-S and {\XMM}/EPIC observations. Exposure times for the {\XMM} observations are reported for the pn, 
  MOS1 and MOS2 detectors respectively, and source net counts refer to the pn detector.}
\begin{threeparttable}
\begin{tabular}{lccccc}
\hline
Obs. ID     						& MJD	& Start time (TT)			& End time (TT)			& Exposure time	& Source net counts\\
							&		& (yyyy/mm/dd hh:mm:ss) 	& (yyyy/mm/dd hh:mm:ss)	& (ks)			& ($\times 10^3$)\\
\hline
14702\tnote{*}					& 56424.55	& 2013/05/12 10:38:50		& 2013/05/12 15:35:56	& 13.7 			& 7.4\\
15040\tnote{**} 				& 56437.63	& 2013/05/25 11:38:37 		& 2013/05/25 18:50:50	& 23.8 			& 3.5\\
14703\tnote{*}					& 56447.48	& 2013/06/04 08:45:16		& 2013/06/04 14:29:15	& 16.8			& 7.6\\
15651\tnote{**} 				& 56448.99	& 2013/06/05 21:32:38 		& 2013/06/06 01:50:11	& 13.8 			& 1.9\\
15654\tnote{**} 				& 56452.25	& 2013/06/09 04:26:16 		& 2013/06/09 07:38:28	& 9.0 			& 1.2\\
14946\tnote{*}					& 56475.41	& 2013/07/02 06:57:56		& 2013/07/02 12:46:18	& 18.2			& 7.1\\
15041						& 56500.36	& 2013/07/27 01:27:17		& 2013/07/27 15:53:25	& 45.4			& 15.7\\
15042						& 56516.25	& 2013/08/11 22:57:58		& 2013/08/12 13:07:47	& 45.7			& 14.4\\
0724210201\tnote{$\dagger$}		& 56535.19	& 2013/08/30 20:30:39		& 2013/08/31 12:28:26	& 55.6/57.2/57.2	& 39.7\\
14945						& 56535.55	& 2013/08/31 10:12:46		& 2013/08/31 16:28:32	& 18.2			& 5.3\\
0700980101\tnote{$\dagger$}		& 56545.37	& 2013/09/10 03:18:13 		& 2013/09/10 14:15:07	& 35.7/37.3/37.3	& 24.9\\
15043						& 56549.30	& 2013/09/14 00:04:52		& 2013/09/14 14:19:20	& 45.4			& 12.5\\
14944						& 56555.42	& 2013/09/20 07:02:56		& 2013/09/20 13:18:10	& 18.2			& 5.0\\
0724210501\tnote{$\dagger$}		& 56558.15	& 2013/09/22 21:33:13		& 2013/09/23 09:26:52	& 41.0/42.6/42.5	& 26.5\\
15044						& 56570.01	& 2013/10/04 17:24:48		& 2013/10/05 07:01:03	& 42.7			& 10.9\\
14943						& 56582.78	& 2013/10/17 15:41:05		& 2013/10/17 21:43:58	& 18.2			& 4.5\\
14704						& 56588.62	& 2013/10/23 08:54:30		& 2013/10/23 20:43:44	& 36.3			& 8.7\\
15045						& 56593.91	& 2013/10/28 14:31:14		& 2013/10/29 05:01:24	& 45.4			& 10.6\\
16508   						& 56709.77	& 2014/02/21 11:37:48  		& 2014/02/22 01:25:55	& 43.4			& 6.8\\
16211 						& 56730.71	& 2014/03/14 10:18:27		& 2014/03/14 23:45:34	& 41.8			& 6.2\\
0690441801\tnote{$\dagger$}		& 56750.72	& 2014/04/03 05:23:24		& 2014/04/04 05:07:01	& 83.5/85.2/85.1	& 34.3\\
16212						& 56751.40	& 2014/04/04 02:26:27		& 2014/04/04 16:49:26	& 45.4			& 6.2\\
16213						& 56775.41	& 2014/04/28 02:45:05		& 2014/04/28 17:13:57	& 45.0			& 5.8\\
16214						& 56797.31	& 2014/05/20 00:19:11		& 2014/05/20 14:49:18	& 45.4			& 5.4\\
16210						& 56811.24	& 2014/06/03 02:59:23		& 2014/06/03 08:40:34	& 17.0			& 1.9\\
16597						& 56842.98	& 2014/07/04 20:48:12		& 2014/07/05 02:21:32	& 16.5			& 1.6\\
16215						& 56855.22	& 2014/07/16 22:43:52 		& 2014/07/17 11:49:38 	& 41.5			& 3.8\\
16216						& 56871.43	& 2014/08/02 03:31:41 		& 2014/08/02 17:09:53	& 42.7 			& 3.6\\
16217						& 56899.43	& 2014/08/30 04:50:12 		& 2014/08/30 15:45:44	& 34.5 			& 2.8\\
0743630201\tnote{$\dagger$}		& 56900.02 	& 2014/08/30 19:37:28 		& 2014/08/31 05:02:43	& 32.0/33.6/33.6	& 9.2	\\
0743630301\tnote{$\dagger$}		& 56901.02	& 2014/08/31 20:40:57		& 2014/09/01 04:09:34	& 25.0/26.6/26.6	& 7.8	\\
0743630401\tnote{$\dagger$}		& 56927.94	& 2014/09/27 17:47:50		& 2014/09/28 03:05:37	& 25.7/32.8/32.8	& 7.7	\\
0743630501\tnote{$\dagger$}		& 56929.12	& 2014/09/28 21:19:11		& 2014/09/29 08:21:11	& 37.8/39.4/39.4	& 11.7\\
\hline
\end{tabular}
\begin{tablenotes}
\item[*] Observations already analysed by Rea et al. (2013a). An additional \AXAF/HRC observation was carried out on 2013 April 29. 
\item[**] {\AXAF} grating observations.
\item[$\dagger$] {\XMM} observations.
\end{tablenotes}
\end{threeparttable}
\label{tab:log}
\end{table*}
%%%%%%%%%%%%%%%%%%%%%%%%%%%%%%%%%%%%%%%%%%%%%%%%%%%%%%%%%%%%%%%%%%%%%%%%%%%%%%%%%%%%%%%%%%

Among the large variety of Galactic neutron stars,\\ \\ magnetars
constitute the most unpredictable class (Mereghetti 2008; Rea 
\& Esposito 2011). They are isolated X-ray pulsars rotating
at relatively long periods ($P\sim2-12$\,s, with spin period 
derivatives $\dot{P}\sim10^{-15}$--$10^{-10}$ s~s$^{-1}$), and
their emission cannot be explained within the commonly accepted 
scenarios for rotation-powered pulsars. In fact, their X-ray luminosity
(typically $L_{\rm X}\sim10^{33}$--$10^{35}$ erg s$^{-1}$) generally 
exceeds the rotational energy loss rate and their temperatures are
often higher than non-magnetic cooling models predict. 
It is now generally recognized that these sources are powered by 
the decay and the instability of their exceptionally high magnetic field 
(up to $B \sim 10^{14}-10^{15}$ G at the star surface), hence the name 
``magnetars'' (Duncan \& Thompson 1992; Thompson \& Duncan 1993; 
Thompson, Lyutikov \& Kulkarni 2002).
Alternative scenarios such as accretion from a fossil disk surrounding the neutron star (Chatterjee, Hernquist \&
Narayan 2000; Alpar 2001) or quark-nova models (Ouyed, Leahy \& Niebergal 2007a,b) have not been ruled out
(see Turolla \& Esposito 2013 for an overview).

%%%%%%%%%%%%%%%%%%%%%%%%%%%%%%%%%%%%%%%%%%%%%%%%%%%%%%%%%%%%%%%%%%%%%%%%%%%%%%%%%%%%%%%%%%
\begin{table*}
%\scriptsize
\caption{Timing solutions. Errors were evaluated at the 1$\sigma$
confidence level, scaling the uncertainties by the value of the rms
($\sqrt{\chi_\nu^2}$) of the respective fit to account for the presence 
of unfitted residuals.}
\begin{tabular}{lcccc}
\hline

Solution 				& Rea et al. (2013a)                 	 &  Kaspi et al. (2014)        	& This work (Solution A)           	& This work (Solution B)         \\
Epoch $T_0$ (MJD)    	                & 56424.5509871         		 &  56513.0                    	& 56513.0                          	& 56710.0                        \\
Validity range (MJD) 	                & 56411.6 -- 56475.3     		 &  56457 -- 56519            	& 56500.1 -- 56594.1               	& 56709.5 -- 56929      	 \\
$P(T_0)$ (s)         		        & 3.7635537(2)         		 	 &  3.76363824(13)             	& 3.76363799(7)                    	& 3.7639772(12)        		 \\
$\dot{P}(T_0)$       	 	        & $6.61(4)\times10^{-12}$ 	         &  $1.385(15)\times10^{-11}$   & $1.360(6)\times10^{-11}$               & $3.27(7)\times10^{-11}$   	  \\
$\ddot{P}$ (s$^{-1}$)	                & $4(3)\times10^{-19}$    		 &  $3.9(6)\times10^{-19}$     	& $3.7(2)\times10^{-19}$                 & $(-1.8\pm0.8)\times10^{-19}$     \\
$\nu(T_0)$ (Hz)      		        & 0.265706368(14)        		 &  0.265700350(9)           	&  0.26570037(5)                    	&  0.26567642(9)        	  \\
$\dot{\nu}(T_0)$ (Hz s$^{-1}$)           &  $-4.67(3)\times10^{-13}$              &  $-9.77(10)\times10^{-13}$    & $-9.60(4)\times10^{-13}$ 	        & $-2.31(5)\times10^{-12}$ 	 \\
$\ddot{\nu}$ (Hz s$^{-2}$)               & $-3(2)\times10^{-20}$ 		 & $-2.7(4)\times10^{-20}$ 	& $-2.6(1)\times10^{-20}$                & $(1.3\pm0.6)\times10^{-20}$     \\
rms residual			        & 0.15 s				 & 51 ms			& 0.396 s				& 1.0 $\mu$Hz			 \\		
$\chi_\nu^2$ (d.o.f.)		        & 0.85 (5)    			         &  1.27 (41) 		        & 6.14 (44)   		                & 0.66 (10)                        \\
\hline
\end{tabular}
\label{tab:timing}
\end{table*}
%%%%%%%%%%%%%%%%%%%%%%%%%%%%%%%%%%%%%%%%%%%%%%%%%%%%%%%%%%%%%%%%%%%%%%%%%%%%%%%%%%%%%%%%%%

%%%%%%%%%%%%%%%%%%%%%%%%%%%%%%%%%%%%%%%%%%%%%%%%%%%%%%%%%%%%%%%%%%%%%%%%%%%%%%%%%%%%%%%%%%
\begin{figure}
\begin{center}
\includegraphics[width=8.5cm]{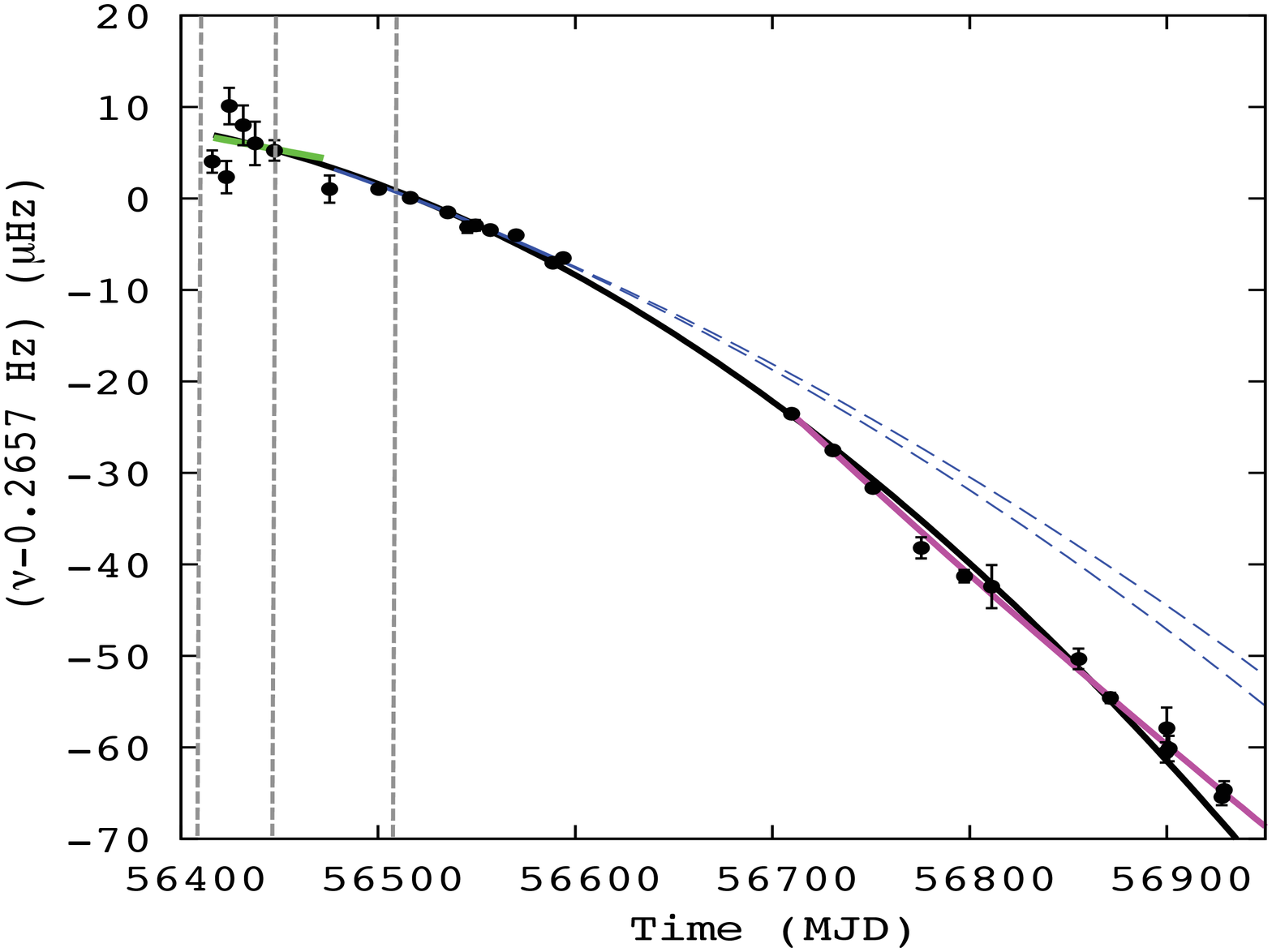}
\includegraphics[width=8.5cm,height=3cm]{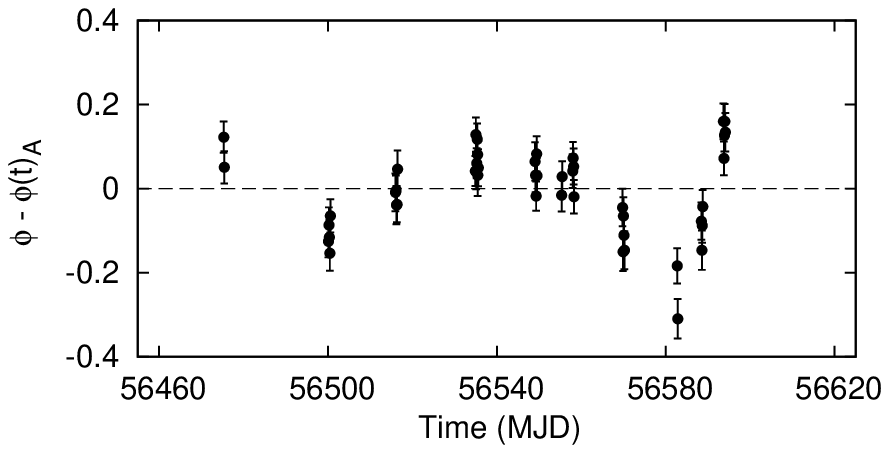}
\includegraphics[width=8.5cm,height=3cm]{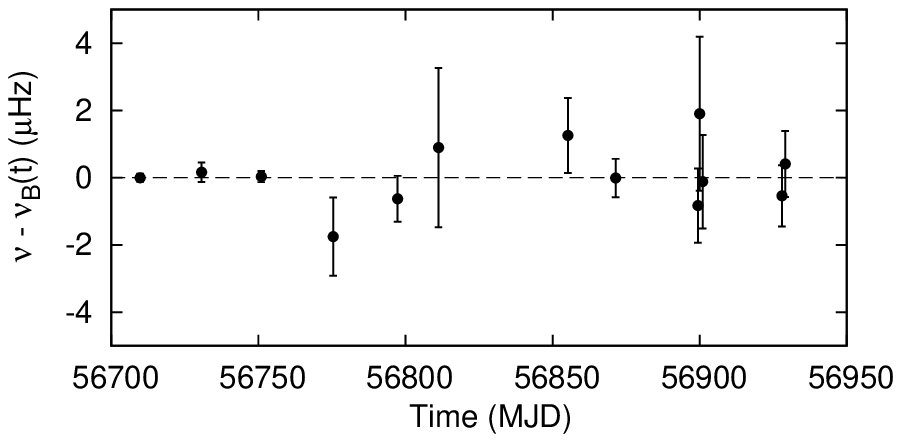}
\end{center}
\caption{{\em Upper panel}: temporal evolution of the spin frequency of \galcen . 
  The solution given by Rea et al. (2013a) is plotted as a green solid line. The blue 
  and magenta solid lines show solutions A and B of this work, respectively. The blue 
  dashed lines are the extrapolation of solution A over the time-span of solution B. 
  The black line represents the fit over the whole time interval covered by observations 
  (see text), while the vertical dashed lines refer to the times of the SGR-like short bursts 
  detected by {\em Swift}/BAT (on 2013 April 25, June 7, and August 5).  {\em Central 
  panel}:  phase residuals with respect to solution A (labelled as $\phi(t)_A$), 
  evaluated over the time validity interval MJD 56500.1 -- 56594.1. {\em Lower panel}: 
  phase residuals with respect to solution B (labelled as $\nu_B(t)$), evaluated over the 
  time validity interval MJD 56709.5 -- 56929.}
\label{fig:timing}
\vskip -0.1truecm
\end{figure}
%%%%%%%%%%%%%%%%%%%%%%%%%%%%%%%%%%%%%%%%%%%%%%%%%%%%%%%%%%%%%%%%%%%%%%%%%%%%%%%%%%%%%%%%%%

The persistent soft X-ray spectrum usually comprises both a thermal 
(blackbody, $kT\sim$ 0.3--0.6 keV) and a non-thermal (power law, 
$\Gamma\sim$ 2--4) components. The former is thought to originate 
from the star surface, whereas the latter likely comes from the reprocessing 
of thermal photons in a twisted magnetosphere through resonant cyclotron 
scattering (Thompson et al. 2002; Nobili, Turolla \& Zane 2008a,b; Rea et al. 
2008; Zane et al. 2009).

In addition to their persistent X-ray emission, magnetars exhibit very
peculiar bursts and flares (with luminosities reaching up to 10$^{46}$ erg
s$^{-1}$ and lasting from milliseconds to several minutes), as well as
large enhancements of the persistent flux (outbursts), which can last
years. These events may be accompanied or triggered by deformations/fractures 
of the neutron star crust (``stellar quakes'') and/or local/global rearrangements 
of the star magnetic field.

In the past decade, extensive study of magnetars in outburst has led to a 
number of unexpected discoveries which have changed our understanding of 
these objects. The detection of typical magnetar-like bursts and a
powerful enhancement of the persistent emission unveiled the existence of three 
low magnetic field ($B <4\times 10^{13}$ G) magnetars (Rea et al. 2010, 2012, 
2014; Scholz et al. 2012; Zhou et al. 2014). Recently, an absorption line at a phase-variable 
energy was discovered in the X-ray spectrum of the low-$B$ magnetar \lowba; 
this, if interpreted in terms of a proton cyclotron feature, provides a direct
estimate of the magnetic field strength close to the neutron star
surface (Tiengo et al. 2013). Finally, a sudden spin-down event, i.e. an anti-glitch, 
was observed for the first time in a magnetar (Archibald et al. 2013).

The discovery of the magnetar \galcen\ dates back to 2013 April 24, 
when the Burst Alert Telescope (BAT) on board the \swift\ satellite detected 
a short hard X-ray burst at a position consistent with that of the supermassive 
black hole at the centre of the Milky Way, Sagittarius A$^*$ (hereafter Sgr~A$^*$). 
Follow-up observations with the \swift\ X-ray Telescope (XRT) enabled characterization of 
the 0.3--10 keV spectrum as an absorbed blackbody (with $kT\sim 1$ keV), and estimate 
a luminosity of $\sim3.9\times 10^{35}$ erg s$^{-1}$ (for an
assumed distance of 8.3 kpc; Kennea et al. 2013a). The following day,
a 94.5\,ks observation performed with the {\em Nuclear Spectroscopic
Telescope Array} ({\em NuSTAR}) revealed 3.76\,s pulsations from the
XRT source (Mori et al. 2013). This measurement was subsequently
confirmed by a 9.8\,ks pointing on April 29 with the High Resolution
Camera (HRC) onboard the {\AXAF} satellite, which was able to single out the 
magnetar counterpart at only 2.4$\pm$0.3 arcsec from Sgr~A$^*$, confirming 
that the new source was actually responsible for the X-ray brightening observed
in the Sgr~A$^*$ region (Rea et al. 2013a). Follow-up observations in the 1.4--20 
GHz band revealed the radio counterpart of the source and detected pulsations 
at the X-ray period (e.g. Eatough et al. 2013a; Shannon \& Johnston 2013). 
The SGR-like bursts, the X-ray spectrum, and the surface dipolar magnetic field 
inferred from the measured spin period and spin-down rate, $B_{\mathrm p}  \sim 
2\times 10^{14}$ G, led to classify this source as a magnetar (Mori et al. 2013; 
Kennea et al. 2013; Rea et al. 2013a).

%%%%%%%%%%%%%%%%%%%%%%%%%%%%%%%%%%%%%%%%%%%%%%%%%%%%%%%%%%%%%%%%%%%%%%%%%%%%%%%%%%%%%%%%%%
\begin{figure*}
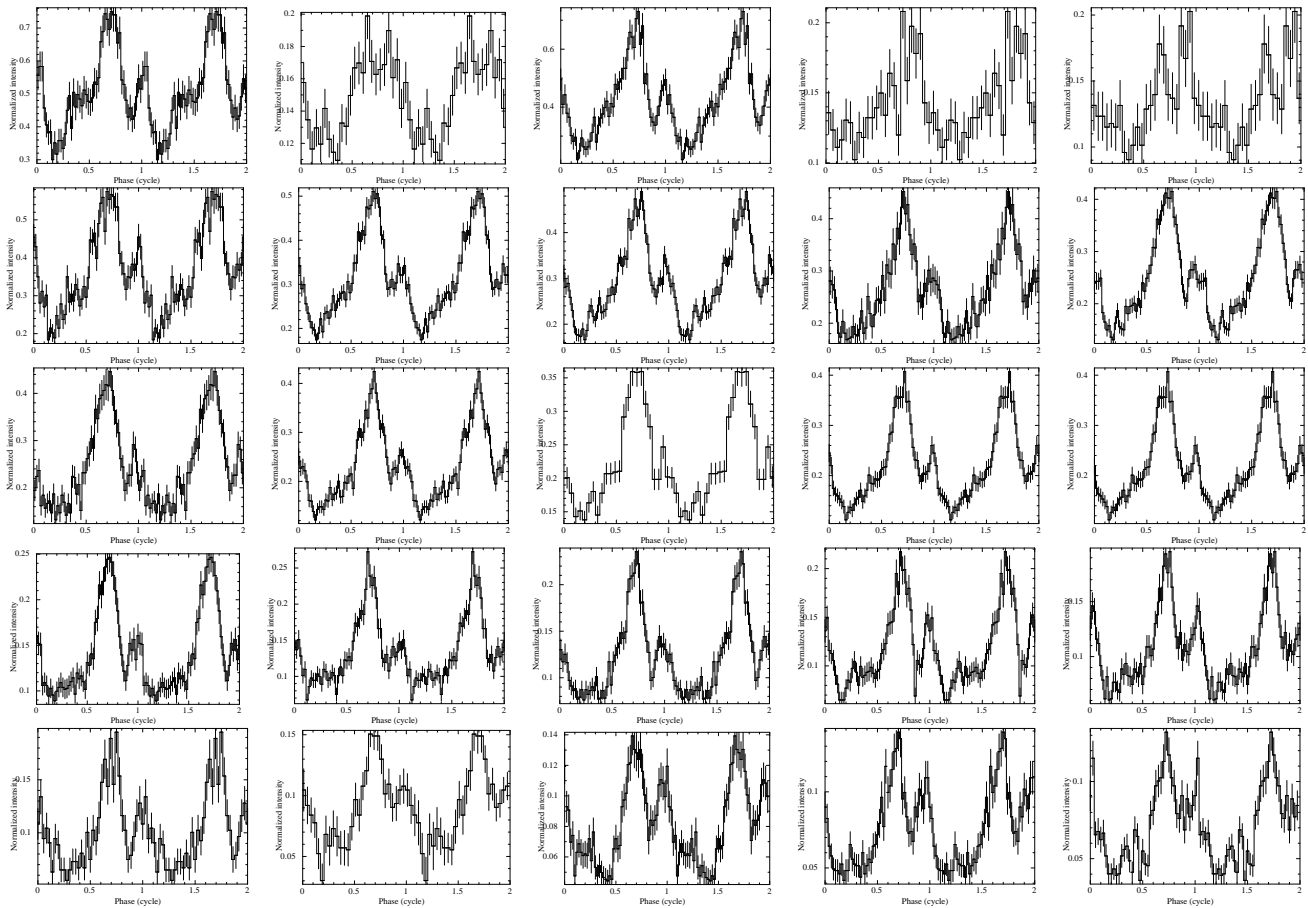

\begin{center}
%\hspace{-1cm}
\includegraphics[width=0.28\columnwidth,angle=-90]{14702lc.ps}
\includegraphics[width=0.28\columnwidth,angle=-90]{15040lc.ps}
\includegraphics[width=0.28\columnwidth,angle=-90]{14703lc.ps}
\includegraphics[width=0.28\columnwidth,angle=-90]{15651lc.ps}
\includegraphics[width=0.28\columnwidth,angle=-90]{15654lc.ps}\\
\includegraphics[width=0.28\columnwidth,angle=-90]{14946lc.ps}
\includegraphics[width=0.28\columnwidth,angle=-90]{15041lc.ps}
\includegraphics[width=0.28\columnwidth,angle=-90]{15042lc.ps}
\includegraphics[width=0.28\columnwidth,angle=-90]{14945lc.ps}
\includegraphics[width=0.28\columnwidth,angle=-90]{15043lc.ps}\\
\includegraphics[width=0.28\columnwidth,angle=-90]{14944lc.ps}
\includegraphics[width=0.28\columnwidth,angle=-90]{15044lc.ps}
\includegraphics[width=0.28\columnwidth,angle=-90]{14943lc.ps}
\includegraphics[width=0.28\columnwidth,angle=-90]{14704lc.ps}
\includegraphics[width=0.28\columnwidth,angle=-90]{15045lc.ps}\\
\includegraphics[width=0.28\columnwidth,angle=-90]{16508lc.ps}
\includegraphics[width=0.28\columnwidth,angle=-90]{16211lc.ps}
\includegraphics[width=0.28\columnwidth,angle=-90]{16212lc.ps}
\includegraphics[width=0.28\columnwidth,angle=-90]{16213lc.ps}
\includegraphics[width=0.28\columnwidth,angle=-90]{16214lc.ps}\\
\includegraphics[width=0.28\columnwidth,angle=-90]{16210lc.ps}
\includegraphics[width=0.28\columnwidth,angle=-90]{16597lc.ps}
\includegraphics[width=0.28\columnwidth,angle=-90]{16215lc.ps}
\includegraphics[width=0.28\columnwidth,angle=-90]{16216lc.ps}
\includegraphics[width=0.28\columnwidth,angle=-90]{16217lc.ps}
\end{center}
\caption{Pulse profiles of \galcen\ obtained from {\AXAF} observations in the 0.3--10 keV energy range. Epoch increases from left to right, top to bottom. Two cycles are shown for clarity.}
\label{fig:pulseprofiles}
\vskip -0.1truecm
\end{figure*}
%%%%%%%%%%%%%%%%%%%%%%%%%%%%%%%%%%%%%%%%%%%%%%%%%%%%%%%%%%%%%%%%%%%%%%%%%%%%%%%%%%%%%%%%%%

\galcen\, holds the record as the closest neutron star to a supermassive black hole 
detected to date. The dispersion measure $DM = 1778\pm3$ cm$^{-3}$\,pc is also 
the highest ever measured for a radio pulsar and is consistent with a
source located within 10 pc of the Galactic Centre.
Furthermore, its neutral hydrogen column density $\nh \sim10^{23}$ cm$^{-2}$ 
is characteristic of a location at the Galactic Centre (Baganoff et al. 2003). 
The angular separation of $2.4 \pm 0.3$ arcsec from Sgr~A$^*$ corresponds 
to a minimum physical separation of $0.09\pm0.02$\,pc (at
a 95 per cent confidence level; Rea et al. 2013a) for an assumed distance of 8.3 kpc
(see e.g. Genzel et al. 2010). Recent observations of the radio counterpart
with the {\em Very Long Baseline Array} ({\em VLBA}) succeeded in 
measuring its transverse velocity of $236\pm11$\,km s$^{-1}$ at position 
angle of $22\pm2$\,deg East of North (Bower et al. 2015). If born within 
1 pc of Sgr~A$^*$, the magnetar has a $\sim$90 per cent probability of 
being in a bound orbit around the black hole, according to the numerical 
simulations of Rea et al. (2013a).

\galcen\, has been monitored intensively in the X-ray and radio bands
since its discovery. Three high-energy bursts were detected from a
position consistent with that of the magnetar on 2013 June 7, August 5
by \swift/BAT, and on September 20 by the {\em INTErnational Gamma-Ray
  Astrophysics Laboratory} ({\INT}) (Barthelmy et al. 2013a,b; Kennea
et al. 2013b,c; Mereghetti et al. 2013).
Kaspi et al. (2014) reported timing and spectral analysis of {\em
NuSTAR} and \swift/XRT data for the first $\sim$4 months of
the magnetar activity (2013 April--August). Interestingly, an
increase in the source spin-down rate by a factor $\sim 2.6$ was
observed, possibly corresponding to the 2013 June burst.
The source has been observed daily with \swift/XRT until 2014 
October, and its 2-10 keV flux has decayed steadily during this time 
interval (Lynch et al. 2015).

Radio observations made possible a value of the rotational measure, 
$RM = 66960 \pm 50$ rad m$^{-2}$, which implies a lower limit 
of $\sim 8$ mG for the strength of the magnetic field in the vicinity of Sgr A$^*$ 
(Eatough et al. 2013b). Observations with the {\em Green Bank Telescope} 
showed that the source experienced a period of relatively stable 8.7-GHz 
flux between 2013 August and 2014 January and then entered a state 
characterized by a higher and more variable flux, until 2014 July (Lynch 
et al. 2015).

In this paper we report on the X-ray long-term monitoring campaign of
\galcen\, covering the first 1.5\,yr of the outburst decay. In Section 2 we
describe the {\AXAF} and {\XMM} observations and the data analysis. In
Section 3 we discuss our results; conclusions follow in Section 4.

\section{Observations and data analysis}

The {\em Chandra X-ray Observatory} observed \galcen\, 26 times
between 2013 April 29 and 2014 August 30. The first observation was
performed with the HRC to have the best spatial accuracy to localize
the source in the crowded region of the Galactic Centre (Rea et
al. 2013a). The remaining observations were performed with the
Advanced CCD Imaging Spectrometer (ACIS; Garmire et al. 2003)
set in faint timed-exposure imaging mode with a 1/8 sub-array (time resolution of
0.4 s), and in three cases with the High Energy Transmission Grating (HETG;
Canizares et al. 2005). The source was positioned on the
back-illuminated S3 chip. Eight observations were carried out by the
{\XMM} satellite using the European Photon Imaging Camera (EPIC), with
the pn (Str\"{u}der et al. 2001) and the two MOS (Turner et al. 2001)
CCD cameras operated in full-frame window mode (time resolution of
73.4 ms and 2.6 s, respectively), with the medium optical blocking
filter in front of them. A log of the X-ray observations is given in Table \ref{tab:log}.

{\AXAF} data were analysed following the standard analysis 
threads\footnote{http://cxc.harvard.edu/ciao/threads/pointlike.} 
with the {\AXAF} Interactive Analysis of Observations software package 
(\textsc{ciao}, version 4.6; Fruscione et al. 2006). {\XMM} data were processed 
using the Science Analysis Software (\textsc{sas}\footnote{http://xmm.esac.esa.int/sas/}, 
version 13.5.0). For both {\AXAF} and {\XMM} data, we adopted the most recent calibration 
files available at the time the data reduction and analysis were performed. 
%Event files for the pn and MOS observations were produced from the raw observation data files with the \textsc{epproc} and \textsc{emproc} \textsc{sas} tools, respectively.

%%%%%%%%%%%%%%%%%%%%%%%%%%%%%%%%%%%%%%%%%%%%%%%%%%%%%%%%%%%%%%%%%%%%%%%%%%%%%%%%%%%%%%%%%%
\begin{figure}
\begin{center}
\includegraphics[width=5.1cm,angle=-90]{pf_all.ps}
\includegraphics[width=9.15cm]{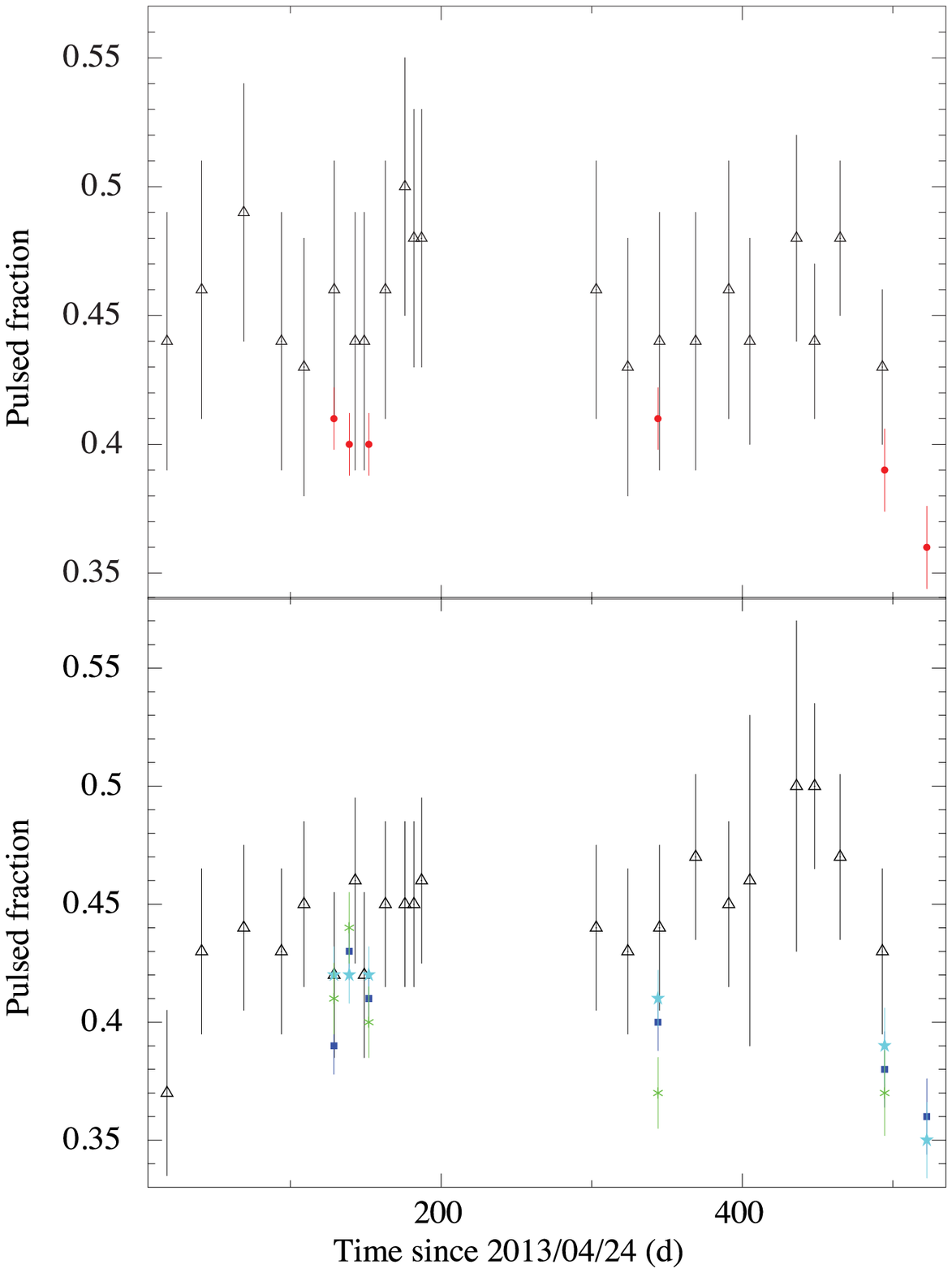}
\end{center}
\caption{Temporal evolution of the pulsed fraction (see text for
  our definition). Uncertainties on the values were obtained by propagating
  the errors on the maximum and minimum count rates. {\em Top panel}: in the 0.3--10 keV band. 
   {\em Central panel}: in the 0.3--3.5 band for the {\AXAF} (black triangles) and {\XMM} (red points)
  observations. {\em Bottom panel}: in the 3.5--10 band for the {\AXAF} observations (black) and in the 
  3.5--5 (blue), 5--6.5 (light blue) and 6.5--10 keV (green) ranges for the {\XMM} observations.} 
\label{fig:pulsefraction}
\vskip -0.1truecm
\end{figure}
%%%%%%%%%%%%%%%%%%%%%%%%%%%%%%%%%%%%%%%%%%%%%%%%%%%%%%%%%%%%%%%%%%%%%%%%%%%%%%%%%%%%%%%%%%

\subsection{Timing analysis}

We extracted all {\AXAF} and {\XMM}/EPIC-pn source counts using a 1.5
and 15-arcsec circles, respectively,  centred on the source position. 
Background counts were extracted using a nearby circular region 
of the same size. We adopted the coordinates reported by Rea et al. 
(2013a), i.e. $\rm RA=17^h45^m40\fs169$, $\rm Dec= -29^\circ00'29\farcs84$ 
(J2000.0), to convert the photon arrival times to Solar System 
barycentre reference frame. The effects of the proper motion 
relative to Sgr~A$^*$ on the source position are negligible on 
the timescales considered for our analysis (best-fit parameters are 
$1.6 < \mu_\alpha < 3.0$ mas yr$^{-1}$ and $5.7 < \mu_\delta < 6.1$ 
mas yr$^{-1}$ at a 95 per cent confidence level; Bower et al. 2015).

To determine a timing solution valid over the time interval
covered by the {\AXAF} and {\XMM} observations (from 2013 April 29 
to 2014 August 30; see Table~\ref{tab:log}), we first considered the timing 
solutions given by Rea et al. (2013a; using {\AXAF} and \swift) and Kaspi 
et al. (2014; using {\em NuSTAR} and \swift). In the overlapping time interval, 
before 2013 June 14 (MJD 56457), both papers report a consistent timing 
solution (see first column in Table~\ref{tab:timing} and green solid line in 
the upper panel of Fig. \ref{fig:timing}).
%but detecting in addition a significant 
%second derivative of the period $\ddot{P} \simeq 1\times10^{-18}$ s$^{-1}$. 
Kaspi et al. (2014) then added more observations covering the interval 
between 2013 June 14 and August 15 (MJD 56457--56519), and 
observed a $\dot{P}$ roughly two times larger than the previous value 
(see Table~\ref{tab:timing}). 
The uncertainties on the Kaspi et al. (2014) solution formally ensure 
unambiguous phase connection until 2013 November 11 (MJD 56607), 
allowing us to extend this phase-coherent analysis with the data reported 
here, and follow the evolution of the pulse phases between 2013 July 27 
and October 28 (MJD 56500--56594; after which we have a gap in our data 
coverage of about 115 days; see Table~\ref{tab:timing}).

In this time interval, we measured the pulse phase at the fundamental frequency 
by dividing our observations in intervals of 10\,ks and using the solution given by 
Kaspi et al. (2014) to determine univocally the number of cycles between the 
various observations. By fitting the measured pulse phases with a cubic function, 
we obtained the solution dubbed A in Table~\ref{tab:timing}, which shows only 
slight deviations with respect to the solution published by Kaspi et al. (2014), but extends 
until 2013 October 28 (MJD 56594). The period evolution implied by solution A is 
plotted with a blue solid line in the upper panel of Fig.~\ref{fig:timing}. Our {\AXAF} 
and {\XMM} observations allow us to confirm the change in the $\dot{P}$, which 
increased by a factor of $\sim$2 around 2013 June (i.e. about two months after the 
onset of the outburst in 2013 April), and remained stable until at least 2013 October 28.

%The impossibility of describing the spin
%evolution over the whole time interval considered with a single
%function is shown by the very large residuals of a solution valid over 
%such an interval (see black solid line in Fig.~\ref{fig:timing}).

%%%%%%%%%%%%%%%%%%%%%%%%%%%%%%%%%%%%%%%%%%%%%%%%%%%%%%%%%%%%%%%%%%%%%%%%%%%%%%%%%%%%%%%%%%
\begin{figure*}
\begin{center}
\hbox{
\includegraphics[height=10cm,width=8cm]{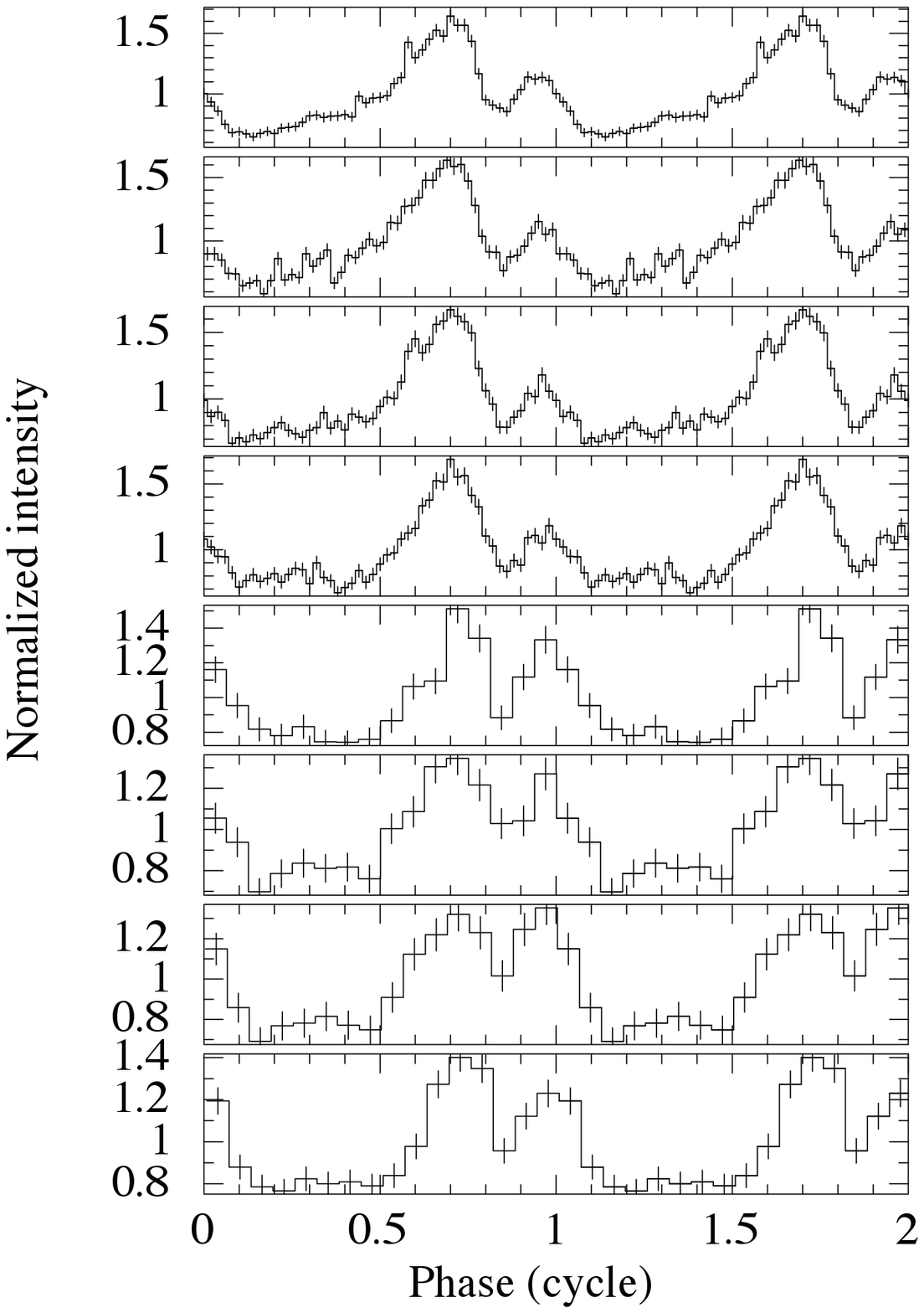}
\hspace{-0.5cm}
\includegraphics[width=8cm,height=9cm]{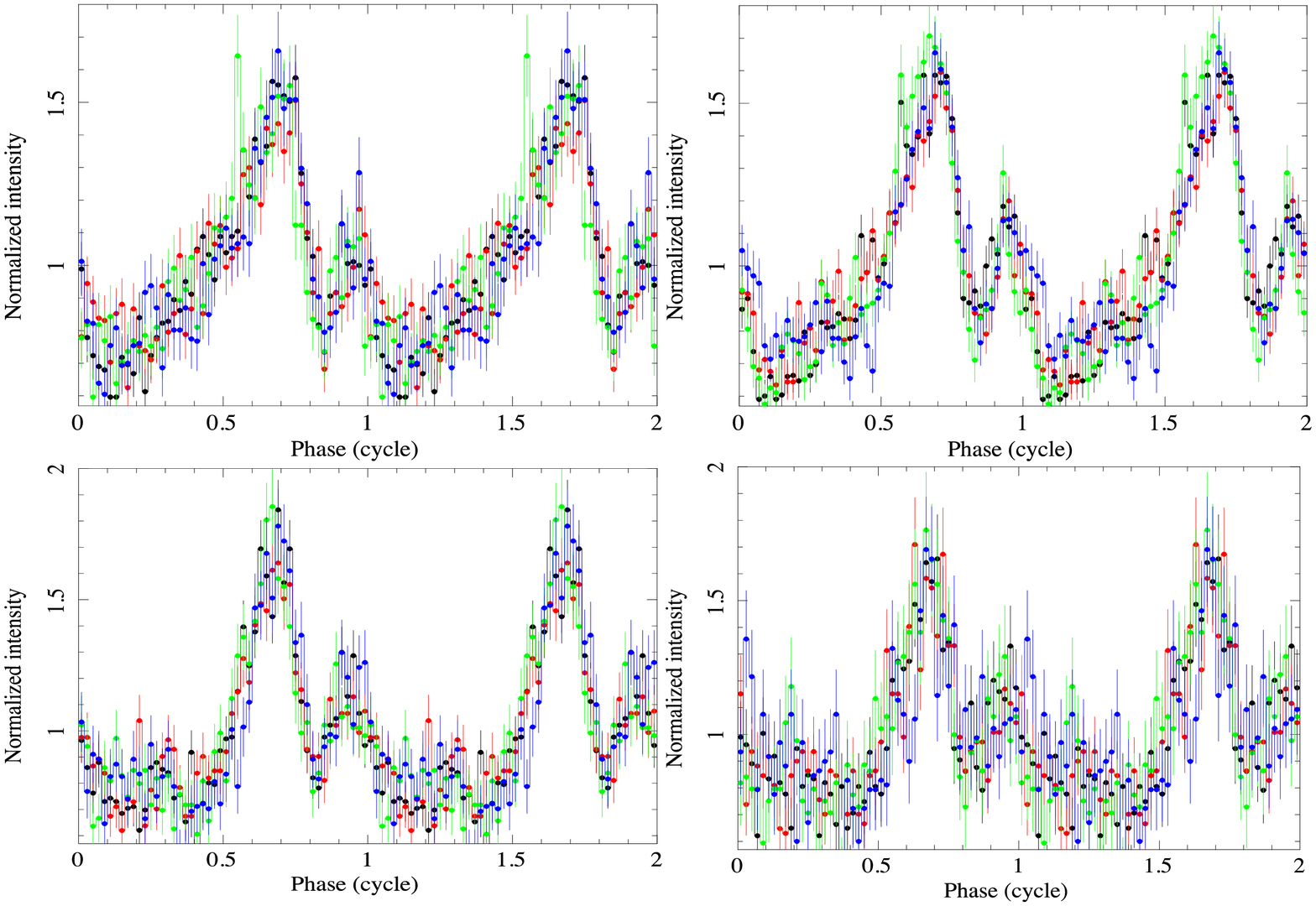}}
\caption{Pulse profiles of \galcen\ obtained from {\XMM}/EPIC-pn observations.  
Two cycles are shown for clarity.  {\em Left}: pulse profiles in the 0.3--10 keV 
energy band. {\em Right}: pulse profiles in the 0.3--3.5, 3.5--5, 5--6.5, 6.5--10 
keV energy bands (from left to right, top to bottom) for the first four observations. 
Black, red, green and blue colors refer to the first, second, third and fourth observation, 
respectively.}
\label{fig:xmmpulsefraction}
\label{fig:xmmpulseprofiles}
\end{center}
\end{figure*}
%%%%%%%%%%%%%%%%%%%%%%%%%%%%%%%%%%%%%%%%%%%%%%%%%%%%%%%%%%%%%%%%%%%%%%%%%%%%%%%%%%%%%%%%%%

Formally, the accuracy of solution A should guarantee that phase
coherence is not lost before 2014 March 3 (MJD 56721), i.e. comprising
the first observation available after the 115 day gap between MJD
56594.1 and MJD 56709.5. However, fitting the phases derived for that
observation with solution A shows large residuals.  These clearly
indicate that solution A is not valid after the gap.  To investigate
this change in the spin evolution of the source, we measured the spin
frequency for all the observations performed after the gap by fitting
with a linear function the phases determined over time intervals with
lengths ranging from 2 to 10 ks, depending on the source flux. The
values for the frequencies we measured in this way after 2014 February
21 (MJD 56709) are much smaller than those predicted by solution A
(see blue dashed line in the upper panel of Fig.~\ref{fig:timing}).

To determine the spin evolution of the source after the 115 day gap in
the observations (i.e. from MJD 56709), we then fitted the values of
the spin frequency with a quadratic function, obtaining the
non-coherent solution B (see Table~\ref{tab:timing}), plotted in the
upper panel of Fig.~\ref{fig:timing} with a magenta solid
line. Unfortunately, this solution is not accurate enough to determine
univocally the number of rotations between the various observations.
Still, the trend followed by the spin frequency after the gap clearly
deviates from that shown before 2013 October 28 via solution A,
indicating a further increase of the spin-down rate. In particular,
the $\dot{P}$ has further increased by a factor of $\sim$2.5, and the
$\ddot{P}$ is smaller than that measured by solution A, even if the
large error prevents us from detecting a change in the sign of the
$\ddot{P}$ at high significance.

The large changes in the timing properties of the source since the
onset of the outburst are also shown by the fact that a quadratic
function gives a poor fit for the spin frequency evolution over the
whole time interval covered by the observations ($\chi^2_\nu=5.04$ for
26 d.o.f.; see black solid line in the upper panel of
Fig.~\ref{fig:timing}).

Summarizing, we derive a phase coherent solution (solution A, see
Table~\ref{tab:timing} and blue solid line in the upper panel of
Fig.~\ref{fig:timing}) that is able to model the pulse phase evolution
before the 115\, day observations gap starting at MJD 56600, and which
is compatible with the solution given by Kaspi et al. (2014) for the
partly overlapping interval MJD 56457 -- 56519. After the observation
gap, solution A is no longer able to provide a good description of
pulse phases, and we are only able to find a solution based on the
analysis of the spin frequency evolution (solution B, see
Table~\ref{tab:timing} and magenta solid line in the upper panel of
Fig. \ref{fig:timing}).

We then use timing solution A (up to MJD 56594.1) and solution B
(from MJD 56709.5 onwards) to fold all background-subtracted and 
exposure-corrected light curves at the neutron star spin period during 
the corresponding observation (see Figs \ref{fig:pulseprofiles} and 
\ref{fig:xmmpulsefraction}). This allows us to extract the temporal evolution 
of the pulsed fraction, defined as PF=[Max -Min]/[Max + Min] (Max and Min 
being the maximum and the minimum count rate of the pulse profile, 
respectively). To investigate possible dependences on energy, we calculate 
the pulsed fractions in the 0.3--3.5 and 3.5--10 keV intervals for the {\AXAF} 
observations and in the 0.3--3.5, 3.5--5, 5--6.5, 6.5--10 keV ranges for the 
{\XMM} observations (see Fig. \ref{fig:pulsefraction}).

%%%%%%%%%%%%%%%%%%%%%%%%%%%%%%%%%%%%%%%%%%%%%%%%%%%%%%%%%%%%%%%%%%%%%%%%%%%%%%%%%%%%%%%%%%%%%%%%%%%%%%%%%%%%%%%%%%%
\begin{table*}
\caption{{\AXAF} spectral fitting results obtained with an absorbed blackbody model ($\chi^2_\nu = 1.00$ for 2282 d.o.f.). The hydrogen column density 
was tied to be the same in all the observations, resulting in $\nh = 1.90(2) \times 10^{23}$ cm$^{-2}$ (photoionization cross-sections from Verner et al. (1996) and 
chemical abundances from Wilms et al. (2000)). The $\alpha$ is a parameter of the \textsc{xspec} pile-up 
model (see Davis 2001 and ``{\em The Chandra ABC Guide to pile-up}''). The pile-up model was not included when fitting the HETG/ACIS-S
spectra (obs. ID: 15040, 15651, 15654). The blackbody radius and luminosity are calculated assuming a source distance of 8.3 kpc (see 
e.g. Genzel et al. 2010). Fluxes and luminosities were calculated after removing the pile-up model. All errors are quoted at a 90 per cent 
confidence level for a single parameter of interest ($\Delta \chi^2 = 2.706$).} 
\begin{tabular}{cccccc}
\hline
Obs. ID 	     		 & $\alpha$				& $kT_{BB}$			& $R_{BB}$			& 1--10 keV absorbed flux			& 1--10 keV luminosity\\ 
		     		 &						& (keV)		   		& (km) 				& (10$^{-12}$ erg cm$^{-2}$ s$^{-1}$)	& (10$^{35}$ erg s$^{-1}$)\\ 
\hline
\vspace{0.08cm}  
14702			  & 0.47(6)					& 0.87(2)				& 2.6$_{-0.1}^{+0.2}$	& 16.5$_{-0.8}^{+1.0}$				& 4.7	(3)				\\ \vspace{0.08cm}  
15040			  & -						& 0.90(2)				& 2.5(1)				& 15.5$_{-1.3}^{+0.03}$				& 4.7	(4)				\\ \vspace{0.08cm}
14703	     		  & 0.47$_{-0.06}^{+0.07}$	& 0.84(2) 				& 2.6(1)  				& 12.7$_{-0.6}^{+0.5}$				& 3.9	(3)				\\ \vspace{0.08cm} 
15651			  & -						& 0.87(3)				& 2.4(2)				& 12.5$_{-0.9}^{+0.07}$				& 3.8	(4)				\\ \vspace{0.08cm}
15654			  & -						& 0.88(4)				& 2.4(2)				& 12.4$_{-0.9}^{+0.05}$				& 3.5	(4)				\\ \vspace{0.08cm}
14946	     		  & 0.43(8)					& 0.82(2) 				& 2.5(1)	 			& 10.5$_{-0.7}^{+0.4}$				& 3.3(3)				\\ \vspace{0.08cm} 
15041	     		  & 0.42$_{-0.05}^{+0.06}$ 	& 0.83(1) 				& 2.22$_{-0.09}^{+0.10}$	& 9.3$_{-0.3}^{+0.2}$				& 2.9	$_{-0.4}^{+0.2}$	\\ \vspace{0.08cm} 
15042	     		  & 0.55(7)					& 0.83(1) 				& 2.14(9) 				& 8.3(3)							& 2.6$_{-0.4}^{+0.2}$	\\ \vspace{0.08cm} 
14945	     		  & 0.4(1)					& 0.85(2) 				& 1.9(1) 				& 7.6$_{-0.4}^{+0.3}$				& 2.3(2)				\\ \vspace{0.08cm} 
15043	     		  & 0.51(8)					& 0.82(1) 				& 2.09$_{-0.09}^{+0.10}$ 	& 7.2$_{-0.3}^{+0.2}$				& 2.4	$_{-0.3}^{+0.2}$	\\ \vspace{0.08cm} 
14944	     		  & 0.6(1)					& 0.84(2) 				& 1.9(1) 				& 7.0(4)							& 2.2$_{-0.3}^{+0.2}$	\\ \vspace{0.08cm} 
15044	     		  & 0.48$_{-0.08}^{+0.09}$	& 0.81(1)				& 2.03$_{-0.09}^{+0.10}$ 	& 6.4(2)							& 2.1$_{-0.3}^{+0.2}$	\\ \vspace{0.08cm} 
14943	     		  & 0.4(1)					& 0.80(2)				& 2.0$_{-0.1}^{+0.2}$ 	& 6.1$_{-0.4}^{+0.2}$				& 2.0(3)				\\ \vspace{0.08cm} 
14704	     		  & 0.5(1)					& 0.80$_{-0.01}^{+0.02}$	& 2.0(1)				& 6.0$_{-0.3}^{+0.2}$				& 2.0	(2)				\\ \vspace{0.08cm} 
15045			  & 0.42(9)					& 0.82$_{-0.01}^{+0.02}$	& 1.88(9)	 			& 5.9$_{-0.2}^{+0.1}$				& 1.9	$_{-0.2}^{+0.1}$	\\ \vspace{0.08cm}
16508	     		  & 0.6(2)					& 0.80$_{-0.01}^{+0.02}$	& 1.65$_{-0.10}^{+0.09}$ 	& 3.8$_{-0.2}^{+0.1}$				& 1.3	$_{-0.2}^{+0.1}$	\\ \vspace{0.08cm}
16211	     		  & 0.3(2)					& 0.79(2) 				& 1.64$_{-0.09}^{+0.10}$ 	& 3.6$_{-0.2}^{+0.1}$				& 1.2	(2)				\\ \vspace{0.08cm}
16212			  & 0.4(2)					& 0.80(2) 				& 1.51$_{-0.09}^{+0.10}$ 	& 3.2(1)							& 1.1(1)				\\ \vspace{0.08cm}
16213			  & 0.3(2)					& 0.79(2) 				& 1.49$_{-0.07}^{+0.08}$ 	& 3.1(1)							& 1.1	(1)				\\ \vspace{0.08cm}
16214			  & 0.4(2)					& 0.79(2) 				& 1.45$_{-0.09}^{+0.10}$ 	& 2.8(1)							& 1.0	(1)				\\ \vspace{0.08cm}
16210			  & 0.4(2)					& 0.82(3) 				& 1.3(1) 				& 2.70(7)							& 0.9(1) 				\\ \vspace{0.08cm}
16597			  & 0.5(2)					& 0.76(3)				& 1.4$_{-0.1}^{+0.2}$	& 2.20$_{-0.05}^{+0.04}$				& 0.8	(1)				\\ \vspace{0.08cm} 
16215			  & 0.4(2) 					& 0.80(2) 	 			& 1.22$_{-0.08}^{+0.09}$ & 2.11(4)							& 0.7	(1)				\\ \vspace{0.08cm}
16216			  & 0.3(2) 					& 0.76(2)				& 1.34$_{-0.09}^{+0.10}$ & 1.91(4)							& 0.7(1)				\\ \vspace{0.08cm}
16217			  & 0.3(2) 					& 0.76(2)				& 1.3(1) 				& 1.80(3)							& 0.67(9)				\\ 
\hline
\end{tabular}
\label{tab:spectralfits}
\end{table*}
%%%%%%%%%%%%%%%%%%%%%%%%%%%%%%%%%%%%%%%%%%%%%%%%%%%%%%%%%%%%%%%%%%%%%%%%%%%%%%%%%%%%%%%%%%%%%%%%%%%%%%%%%%%%%%%%%%%

\subsection{Spectral analysis of {\AXAF} observations}

For all the {\AXAF} observations, we extracted the source counts from a
1.5-arcsec radius circular region centred on \galcen. This corresponds to 
an encircled energy fraction of $\sim85$ per cent of the {\AXAF}
point spread function (PSF) at 4.5 keV. A
larger radius would have included too many counts from the Sgr A$^*$ 
PSF, overestimating the flux of \galcen\, with only a marginal increase 
of the encircled energy fraction (less than $\sim5$ per cent). We extracted
the background counts using three different regions: an 
annulus (inner and outer radius of 14 and 20 arcsec, respectively), 
four 2-arcsec radius circles arranged in a square centred on the source, 
or a 1.5-arcsec radius circle centred on the source position in an archival 
{\AXAF}/ACIS-S observation (i.e. when the magnetar was still in quiescence). 
For grating observations we considered instead a circle of radius 10
arcsec as far as possible from the grating arms but including part of
the diffuse emission present in the Galactic Centre.

For `non-grating' observations, we created the source and background spectra, the 
associated redistribution matrix files and ancillary response files using the 
\textsc{specextract} tool\footnote{Ancillary response 
files are automatically corrected to account for continuous degradation in the 
ACIS CCD quantum efficiency.}. For the three grating observations, we analyzed 
only data obtained with the High Energy Grating (0.8--8 keV). In all cases 
\galcen\ was offset from the zeroth-order aim point, which was centered on the 
nominal Sgr A$^*$ coordinates ($\rm RA =17^h45^m40\fs00$, $\rm Dec= -29^\circ00'28\farcs1$ (J2000.0)). 
We extracted zeroth-order spectra with the \textsc{tgextract} tool and generated
redistribution matrices and ancillary response files using \textsc{mkgrmf} 
and \textsc{fullgarf}, respectively.

%%%%%%%%%%%%%%%%%%%%%%%%%%%%%%%%%%%%%%%%%%%%%%%%%%%%%%%%%%%%%%%%%%%%%%%%%%%%%%%%%%%%%%%%%%
\begin{figure*}
\begin{center}
\includegraphics[width=10cm,angle=-90]{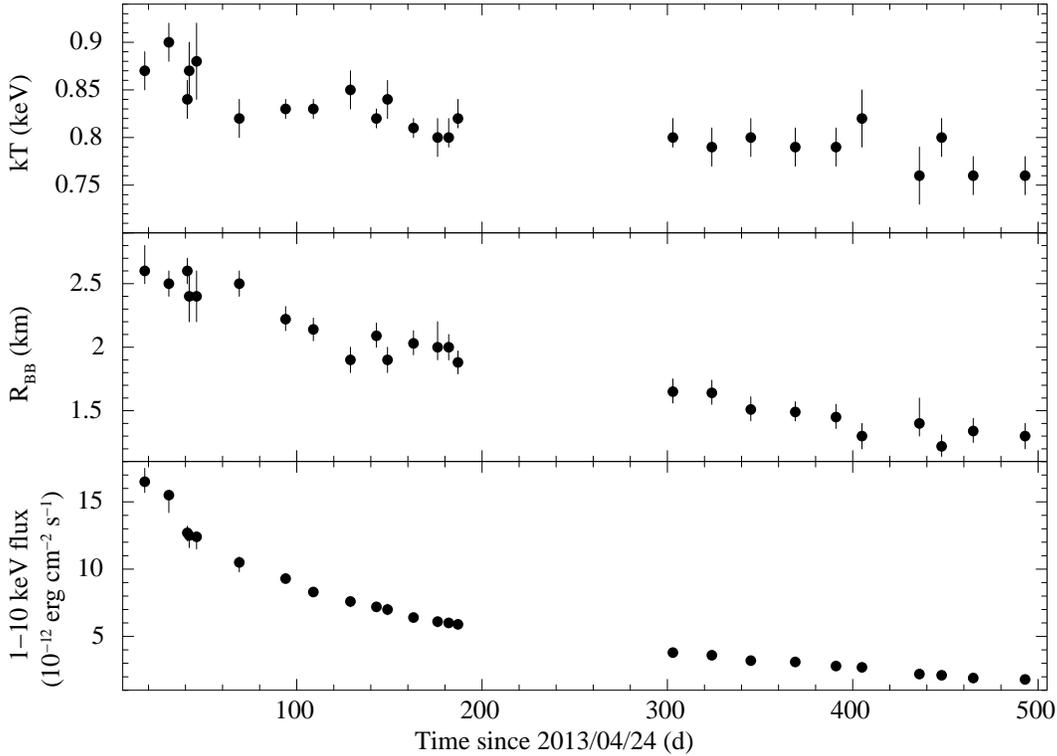}
\end{center}
\caption{Temporal evolution of the spectral parameters for the blackbody model and of the 1--10 keV absorbed flux of \galcen\ from {\AXAF} observations.}
\label{fig:parameters}
\vskip -0.1truecm
\end{figure*}
%%%%%%%%%%%%%%%%%%%%%%%%%%%%%%%%%%%%%%%%%%%%%%%%%%%%%%%%%%%%%%%%%%%%%%%%%%%%%%%%%%%%%%%%%%

We grouped background-subtracted spectra to have at least 50 counts per 
energy bin, and fitted in the 0.3--8 keV energy band (0.8--8 keV for grating 
observations) with the \textsc{xspec}\footnote{http://heasarc.gsfc.nasa.gov/xanadu/xspec/} 
spectral fitting package (version 12.8.1g; Arnaud 1996), using the $\chi^2$ statistics. 
The photoelectric absorption was described through the \textsc{tbabs} 
model with photoionization cross-sections from Verner et al. (1996) and 
chemical abundances from Wilms, Allen \& McCray (2000). The small 
{\AXAF} PSF ensures a negligible impact of the background at low energies
and allows us to better constrain the value of the hydrogen column 
density towards the source.

We estimated the impact of photon pile-up in the non-grating observations 
by fitting all the spectra individually. Given the pile-up fraction (up to $\sim30$ 
per cent for the first observation as determined with Web\textsc{pimms}, 
version 4.7), we decided to correct for this effect using the pile-up model 
of Davis (2001), as implemented in \textsc{xspec}. According to
`{\em The Chandra ABC Guide to pile-up}',\footnote{http://cxc.harvard.edu/ciao/download/doc/pile-up$_-$abc.pdf}
%the following parameters were left frozen: the maximum number of
%photons considered for pile-up in a single frame, the grade correction
%for single photon detection, and the number of independent $3 \times 3$ 
%pixel pile-up islands in the source extraction region. 
the only parameters allowed to vary were the grade-migration parameter
($\alpha$), and the fraction of events in the source extraction region
within the central, piled up, portion of the PSF. 
%The default value of
%the latter parameter is 95 per cent, which means that 95 per cent of
%the spectrum is subject to pile-up, and 5 per cent is left unpiled. The 
%fraction subject to pile-up correction is expected to be
%larger for small extraction regions, thus we left this parameter free
%to vary in the spectral fitting process. However, this parameter was
%found to be consistent with staying constant within the errors at
%$\sim95$ per cent for all the observations. 
Including this component in the spectral modelling, the fits quality 
and the shape of the residuals improve substantially especially for the 
spectra of the first 12 observations (from obs ID 14702 to 15045), when the 
flux is larger. We then compared our results over the three different background 
extraction methods (see above) and found no significant differences in the parameters, 
implying that our reported results do not depend significantly on the exact
location of the selected background region. 

We fitted all non-grating spectra together, adopting four different models: 
a blackbody, a power law, the sum of two blackbodies, and a blackbody 
plus a power law. For all the models, we left all parameters free to vary. 
However, the hydrogen column density was found to be consistent 
with being constant within the errors\footnote{Here, and in the following, 
uncertainties are quoted at the 90 per cent confidence level, unless 
otherwise noted.} among all observations and thus was tied to be the same.
We then checked that the inclusion of the pile-up model in the joint fits did not alter
the spectral parameters for the last 10 observations (from obs ID 16508 onwards), 
when the flux is lower, by fitting the corresponding spectra individually without the pile-up 
component. The values for the parameters are found to be consistent with being the 
same in all cases.

%%%%%%%%%%%%%%%%%%%%%%%%%%%%%%%%%%%%%%%%%%%%%%%%%%%%%%%%%%%%%%%%%%%%%%%%%%%%%%%%%%%%%%%%%%
\begin{table*}
\caption{{\XMM}/EPIC-pn spectral fitting results obtained with an absorbed blackbody plus power law model ($\chi^2_\nu =1.13$ for 624 d.o.f.) and an absorbed 3D resonant 
cyclotron scattering model ($\chi^2_\nu =1.14$ for 624 d.o.f.). $\beta_{bulk}$ denotes the bulk motion velocity of the charges in the magnetosphere and $\Delta\phi$ is the twist 
angle. For both models the hydrogen column density was tied to be the same in all the observations, yielding $\nh =1.86_{-0.08}^{+0.10} \times 10^{23}$ cm$^{-2}$ for the former 
and $\nh=1.86_{-0.03}^{+0.05} \times 10^{23}$ cm$^{-2}$ for the latter (photoionization cross-sections from Verner et al. (1996) and chemical abundances from Wilms et al. 
(2000)). The blackbody emitting radius and luminosity are calculated assuming a source distance of 8.3 kpc (see e.g. Genzel et al. 2010). Fluxes were determined with the 
\textsc{cflux} model in \textsc{xspec}. All errors are quoted at a 90 per cent confidence level for a single parameter of interest ($\Delta \chi^2 = 2.706$).}
\begin{tabular}{ccccccc}
\hline \vspace{0.2cm}
BB+PL	          	   &						&					&						&						&								&\\							
Obs. ID 	     		   & $kT_{\rm BB}$			& $R_{\rm BB}$		& $\Gamma$ 		 		& PL norm 		 		& 1--10 keV BB/PL abs flux			& 1--10 keV BB/PL luminosity\\ 
		     	  	   & (keV)		   			& (km) 				&				 		& (10$^{-3}$)  				& (10$^{-12}$ erg cm$^{-2}$ s$^{-1}$) 	& (10$^{35}$ erg s$^{-1}$)\\ 
\hline  
\vspace{0.15cm}  
0724210201		   & 0.79(3)				& 1.9(2)				& 2.3$_{-0.7}^{+0.5}$		& 4.5$_{-3.5}^{+8.9}$		& 5.0(1) /  3.3(2)					& 1.7$_{-0.3}^{+0.2}$ / 1.0$_{-0.3}^{+0.4}$	\\ \vspace{0.15cm}  
0700980101		   & 0.78(3)				& 2.1(2)				& 1.7$_{-1.3}^{+0.8}$		& $<6.8$					& 5.7(1) / 2.2(2)						& 2.0(3) / 0.5(3)							\\ \vspace{0.15cm}
0724210501		   & 0.79(4)				& 2.1$_{-0.2}^{+0.3}$	& 2.3$_{-0.6}^{+0.5}$		& $<4.5$ 					& 5.8(1) / 1.8(2) 					& 2.0$_{-0.2}^{+0.1}$ / 0.3$_{-0.2}^{+0.3}$	\\ \vspace{0.15cm}
0690441801	     	   & 0.72$_{-0.04}^{+0.03}$ 	& 1.6(3)  				& 2.6$_{-0.8}^{+0.5}$		& 4.5$_{-3.8}^{+8.8}$		& 1.9(1) / 2.1$_{-0.2}^{+0.1}$ 			& 0.8(2) / 0.8(3)							\\ \vspace{0.15cm} 	
0743630201-301	   & 0.71(6)				& 1.3(3)				& 2.1$_{-1.4}^{+0.7}$		& 1.6$_{-1.5}^{+6.1}$		& 1.2(1) / 1.7(2)						& 0.5(2) / 0.4(3)							\\ \vspace{0.15cm}
0743630401-501	   & 0.67$_{-0.07}^{+0.10}$	& 1.2(5)				& 2.0$_{-0.7}^{+0.4}$		& 6.3$_{-4.9}^{+9.7}$		& 0.7(1) / 2.0(4)						& 0.3(2) / $<0.5$						\\ 
\hline \hline \vspace{0.2cm}
NTZ			   	   &						&					&						&						&								&\\
Obs. ID 	     		   & $kT$ 					& $\beta_{bulk}$ 		& $\Delta\phi$		 		& NTZ norm				& 1--10 keV abs flux					& 1--10 keV luminosity\\ 
		     		   & (keV)		   			&  					& (rad) 			 		& (10$^{-1}$)				& (10$^{-12}$ erg cm$^{-2}$ s$^{-1}$) 	& (10$^{35}$ erg s$^{-1}$) \\ 
\hline  
\vspace{0.15cm}  
0724210201		   & 0.85(2)				& 0.72$_{-0.40}^{+0.09}$ 	& 0.40$_{-0.24}^{+0.04}$		& 1.62$_{-0.12}^{+0.07}$		& 8.3(1)  							& 2.5(2) 	 \\ \vspace{0.15cm}  
0700980101		   & 0.85$_{-0.03}^{+0.02}$	& 0.70$_{-0.34}^{+0.04}$	& 0.40$_{-0.23}^{+0.03}$		& 1.58$_{-0.11}^{+0.14}$		& 7.9(1)							& 2.3(2)	\\ \vspace{0.15cm}
0724210501		   & 0.84(2)				& 0.6(2)				& 0.41$_{-0.25}^{+0.02}$		& 1.5(1)	  				& 7.6(1)			  				& 2.3(3)	 \\ \vspace{0.15cm}
0690441801	     	   & 0.77$_{-0.06}^{+0.04}$ 	& 0.5$_{-0.2}^{+0.3}$  	& 0.42$_{-0.25}^{+0.06}$		& 0.94$_{-0.07}^{+0.10}$		& 4.0(1)  							& 1.3(2)	 \\ \vspace{0.15cm} 	
0743630201-301	   & 0.76$_{-0.10}^{+0.07}$	& $>0.2$				& 0.43$_{-0.03}^{+0.64}$		& 0.61$_{-0.06}^{+0.09}$		& 2.9(1)							& 0.9(3)	\\ \vspace{0.15cm}
0743630401-501	   & 0.65$_{-0.24}^{+0.07}$	& 0.32$_{-0.09}^{+0.11}$	& 0.60$_{-0.17}^{+0.78}$		& 0.68$_{-0.07}^{+0.27}$		& 2.7(1)							& 0.9(3)	\\ 
\hline \hline
\end{tabular}
\label{tab:XMMspectralfits}
\end{table*}
%%%%%%%%%%%%%%%%%%%%%%%%%%%%%%%%%%%%%%%%%%%%%%%%%%%%%%%%%%%%%%%%%%%%%%%%%%%%%%%%%%%%%%%%%%

A fit with an absorbed blackbody model yields $\chi^2_\nu = 1.00$ 
for 2282 degrees of freedom (d.o.f.), with a hydrogen column density 
$\nh = 1.90(2) \times 10^{23}$ cm$^{-2}$, temperature in the 0.76--0.90 
keV range, and emitting radius in the 1.2--2.5 km interval. When an 
absorbed power law model is used ($\chi^2_\nu =1.05$ for 2282 d.o.f.), 
the photon index is within the range 4.2--4.9, much larger than what is 
usually observed for this class of sources (see Mereghetti 2008; Rea \& 
Esposito 2011 for reviews). Moreover, a larger absorption 
value is obtained ($\nh \sim 3 \times 10^{23}$ cm$^{-2}$). The large values for the photon 
index and the absorption are likely not intrinsic to the source, but rather an artifact 
of the fitting process which tends to increase the absorption to compensate for the large 
flux at low energies defined by the power law. The addition of a second component 
to the blackbody, i.e.  another blackbody or a power law, is not statistically required 
($\chi^2_\nu = 1.00$ for 2238 d.o.f. in both cases). We thus conclude that a single 
absorbed blackbody provides the best modelling of the source spectrum in the 0.3--8 keV 
energy range (see Table \ref{tab:spectralfits}). 

Taking the absorbed blackbody as a baseline, we tried to model all the
spectra tying either the radius or the temperature to be the same for
all spectra.  We found $\chi^2_\nu = 1.38$ for 2303 d.o.f. when the
radii are tied, with $\nh = 1.94(2) \times 10^{23}$ cm$^{-2}$, $R_{\rm
  BB}=1.99_{-0.05}^{+0.06}$ km and temperatures in the 0.66--0.97 keV
range. We found instead $\chi^2_\nu = 1.04$ for 2303 d.o.f. when the
temperatures are tied, with $\nh = 1.89(2) \times 10^{23}$ cm$^{-2}$,
$kT_{BB}=0.815(7)$ keV and radii spanning from $\sim$1.1 to $\sim$3
km. The goodness of fit of the latter model improves considerably if
the temperatures are left free to vary as well ($F$-test probability
of $\sim 2 \times 10^{-17}$; fitting the temperature evolution with a
constant yields a poor $\chi^2_\nu =2.8$ for 24 d.o.f. in this case).
We conclude that both the temperature and the size of the blackbody
emitting region are varying.  Zeroth-order spectral data of the three
grating observations were fitted together and indipendently with this model,
without including the pile-up component and fixing $\nh$ to that
obtained in non-grating fit: $1.9 \times 10^{23}$ cm$^{-2}$ (see Table
\ref{tab:spectralfits} and Fig. \ref{fig:parameters}).

%%%%%%%%%%%%%%%%%%%%%%%%%%%%%%%%%%%%%%%%%%%%%%%%%%%%%%%%%%%%%%%%%%%%%%%%%%%%%%%%%%%%%%%%%%
\begin{figure*}
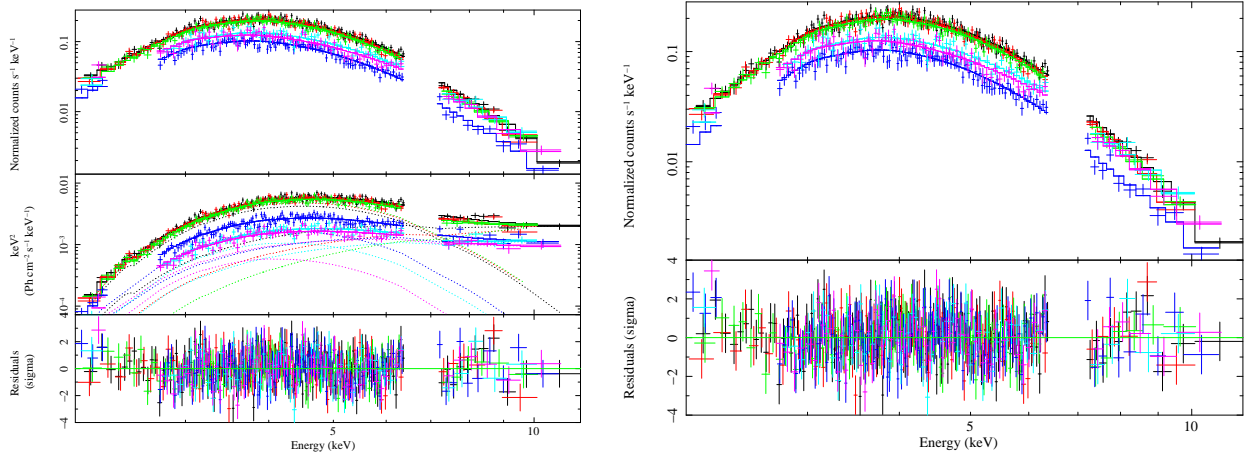

\begin{center}
\includegraphics[width=6cm,angle=-90]{bbpl.ps}
\includegraphics[width=6cm,angle=-90]{ntz.ps}
\caption{Results of the phase-averaged spectral analysis for the {\XMM}/EPIC-pn observations of \galcen. 
  %Black, red and green colors refer to the first, second and third observation, respectively. 
  {\em Left panel}: source spectra fitted together with an absorbed blackbody plus power law model in the 
  2--12 keV range and after removal of the Fe XXV and S XV lines (see text). $E^2\times f(E)$ unfolded spectra 
  together with the contributions of the two additive components and residuals (in units of standard deviations) 
  are also shown.  {\em Right panel}: source spectra fitted together with an absorbed 3-D resonant cyclotron 
  scattering model in the 2--12 keV range and after removal of the Fe XXV and S XV lines (see text). Residuals 
  (in units of standard deviations) are also shown.}
\label{fig:pnspectra}
\end{center}
\end{figure*}
%%%%%%%%%%%%%%%%%%%%%%%%%%%%%%%%%%%%%%%%%%%%%%%%%%%%%%%%%%%%%%%%%%%%%%%%%%%%%%%%%%%%%%%%%%

\subsection{Spectral analysis of {\XMM} observations}

For all the {\XMM} observations, we extracted the source counts from a circular region of radius 15 arcsec centred on the source PSF, and the 
background counts through the same circle at the same position in an archival (2011) {\XMM} observation of the Galactic Centre (obs. ID 
0694640301), when the magnetar was not detected and no transient events were identified within the source PSF.
We built the light curves for the source and background event files to visually inspect and filter for high particle background flaring in the selected 
regions. We checked for the potential impact of pile-up using the \textsc{epatplot} task of \textsc{sas}: the observed pattern distributions for 
both single and double events are consistent with the expected ones (at a 1$\sigma$ confidence level) for all the three cameras, proving that 
the {\XMM} data are unaffected by pile-up. 

We restricted our spectral analysis to photons having \textsc{flag} = 0 and \textsc{pattern} $\leq4(12)$ 
for the pn (MOSs) data and created spectral redistribution matrices and ancillary response files.
We co-added the spectral files of consecutive observations (obs. ID 0743630201-301 and 0743630401-501; see Table \ref{tab:log}) to 
improve the fit statistics and reduce the background contamination.
We then grouped the source spectral channels to have at least 200 counts per bin and fitted the spectra in the 2--12 keV range, given the high 
background contamination within the source PSF at lower energies. The spectral data extracted from the two MOS cameras gave values for the 
parameters and fluxes consistent with those obtained from the pn camera. To minimize the systematic errors introduced when using different 
instruments, we considered only the pn data, which provide the spectra with the highest statistics.

Due to the large PSF of {\XMM}, it is not possible to completely remove 
the contamination of both the Galactic Centre soft X-ray diffuse emission 
and the emission lines from the supernova remnant Sgr A East, including 
in particular the iron line (Fe XXV; rest energy of 6.7 keV) and the sulfur line 
(S XV; rest energy of 2.46 keV) (see e.g. Maeda et al. 2002; Sakano et al. 
2004; Ponti et al. 2010, 2013; Heard \& Warwick 2013). These features were 
clearly visible especially in the spectra of the last observations, when the 
flux is lower, and prevented us from obtaining a good spectral modelling in 
\textsc{xspec}. We thus decided to discard the energy interval comprising the 
Fe XXV line (6.4--7.1 keV) for all the spectra, as well as that associated with the S XV
line (2.3--2.7 keV) for the spectrum of the last observations
(obs. ID 0690441801, 0743630201-301, 0743630401-501), 
involving a loss of $\sim9$ per cent in the total number of spectral bins.

Based on the results of the {\AXAF} spectral analysis, we fitted the
data first with an absorbed blackbody model. The hydrogen column 
density was consistent with being constant at a 90 per cent confidence 
level among all observations and was tied to be the same in the spectral fitting.
We obtained $\chi^2_\nu =2.2$ for 636 d.o.f., with large residuals at high energies. 
The latter disappear if an absorbed power law component is added and 
the fit improves considerably ($\chi^2_\nu =1.13$ for 624 d.o.f.; see left 
panel of Fig. \ref{fig:pnspectra}). A fit with a two-blackbody model is
statistical acceptable as well ($\chi^2_\nu =1.13$ for 624 d.o.f.)
and yields temperatures of $\sim2-4$ keV and emitting
radii of $\sim0.04-0.12$ km for the second blackbody. However, this
model would be physically hard to justify, since it is unlikely
that these large temperatures can be maintained on a neutron star 
surface for such a long time. As an alternative to these fits, we applied
a 3D resonant cyclotron scattering model (NTZ: Nobili et al. 2008a,b;
Zane et al. 2009), obtaining $\chi^2_\nu =1.14$ for 624 d.o.f.  (see
right panel of Fig. \ref{fig:pnspectra}). The hydrogen column densities 
and fluxes inferred both from the BB+PL and the NTZ models are consistent 
with each other within the errors (see Table \ref{tab:XMMspectralfits}).
To test the robustness of our results, we compared the inferred parameters
with those derived by fitting the spectra without filtering for the spectral 
channels and applying the \textsc{varabs} model for the absorption, which 
allows the chemical abundances of different elements to vary (only the 
sulfur and iron abundances were allowed to vary for the present purpose). 
We found consistent values over the two methods.

%Absorbed fluxes determined from {\XMM} observations are slightly overestimated compared to the {\AXAF} observations regarding the same epoch, 
%possibly resulting from the contamination of the X-ray diffuse emission in the {\XMM} data, which is unavoidable due to much larger aperture of the 
%EPIC cameras compared to the aperture of the ACIS instrument.

We conclude that both models successfully reproduce the soft X-ray part of 
the \galcen\ spectra up to $\sim$12 keV, implying that, similar
to other magnetars, the reprocessing of the thermal emission by a dense, twisted 
magnetosphere produces a non-thermal component. The power law detected by \XMM\, 
is consistent with that observed by {\em NuSTAR} (Kaspi et al. 2014), and its very 
low contribution below 8\,keV is consistent with its non-detection in our {\AXAF} data.

\subsection{Pulse phase-resolved spectral (PPS) analysis}

To search for spectral variability as a function of rotational phase and time, we first extracted 
all the spectra of the {\AXAF} observations selecting three pulse phase intervals 
(see Fig.~\ref{fig:pulseprofiles}): peak ($\phi$=0.5--0.9),  minimum ($\phi$=0.2--0.5), 
and secondary peak ($\phi$=0.9--1.2).  We adopted the same extraction regions 
and performed the same data analysis as for the phase-averaged spectroscopy.

For each of the three different phase intervals, we fitted the spectra of all {\AXAF} 
observations jointly in the 0.3--8 keV energy band with an absorbed blackbody 
model and tying the hydrogen column density to be the same in all the observations 
(the pile-up model was included). Since the values of the column density are consistent
with being the same at a 90 per cent confidence level ($1.90(4) \times 10^{23}$ cm$^{-2}$, 
$1.82(4) \times 10^{23}$ cm$^{-2}$, and $1.83_{-0.04}^{+0.05} \times 10^{23}$ cm$^{-2}$ 
for the peak, the secondary peak and the minimum, respectively), we fixed $\nh$ to 
$1.9 \times 10^{23}$ cm$^{-2}$, i.e. to the best-fit value determined with the phase-averaged 
spectroscopy (see Table \ref{tab:spectralfits}). We obtained a good fit in all cases, with 
$\chi^2_\nu =1.04$ for 1005 d.o.f. for the peak, $\chi^2_\nu =1.10$ for 635 d.o.f. for the 
secondary peak, and $\chi^2_\nu = 0.99$ for 713 d.o.f. for the pulse minimum.
The fit residuals were not optimal for energies $\gtrsim6-7$ keV for 
the peak spectra, due to the larger pile-up fraction. 
We extracted the source counts excluding the central
piled up photons (within a radial distance of 0.7 arcsec from the
source position), and repeated the analysis for the peak spectra: the residuals
are now well shaped, and the inferred values for
the spectral parameters did not change significantly.

\begin{figure}
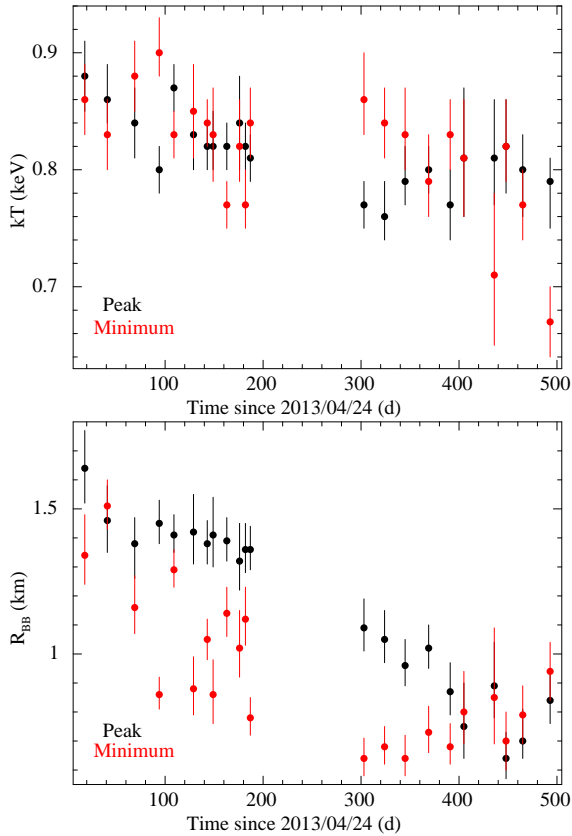

\begin{center}
\includegraphics[width=5.5cm,angle=-90]{chandra_pps_temperature.ps}
\includegraphics[width=5.5cm,angle=-90]{chandra_pps_radius.ps}
\end{center}
\caption{Evolution of the blackbody temperatures ({\em left panel}) and radii ({\em right panel}) for the peak (black points) and the minimum (red points) of the
pulse profile for the {\AXAF} observations.}
\label{fig:PPS}
\vskip -0.1truecm
\end{figure}

The temporal evolution of the blackbody
temperature and radius for both the peak and the pulse minimum are
shown in Fig. \ref{fig:PPS}. No particular trend is observed for the inferred 
temperatures, whereas the size of the emitting region is systematically
lower for the pulse minimum. This is consistent with a viewing geometry 
that allows us to observe the hot spot responsible for the thermal emission almost 
entirely at the peak of the pulse profile, and only for a small fraction at the minimum of the pulsation.

%%%%%%%%%%%%%%%%%%%%%%%%%%%%%%%%%%%%%%%%%%%%%%%%%%%%%%%%%%%%%%%%%%%%%%%%%%%%%%%%%%%%%%%%%%
%%%%%%%%%%%%%%%%%%%%%%%%%%%%%%%%%%%%%%%%%%%%%%%%%%%%%%%%%%%%%%%%%%%%%%%%%%%%%%%%%%%%%%%%%%

\begin{figure*}
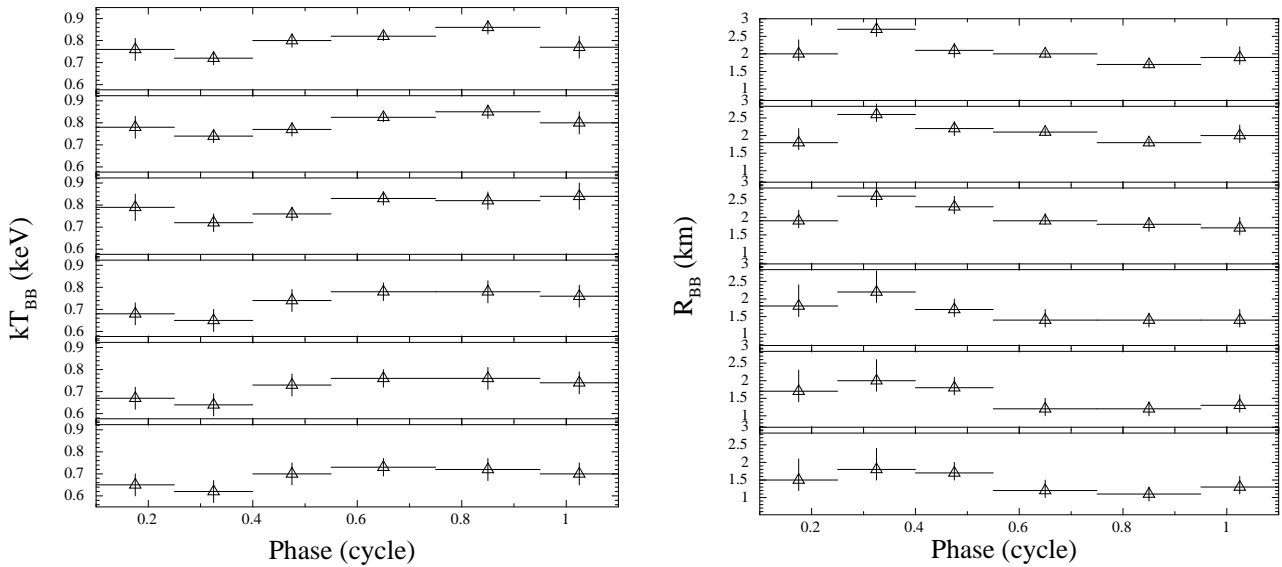

\begin{center}
\includegraphics[width=7.5cm,angle=-90]{xmm_pps_temperature.ps}
\includegraphics[width=7.5cm,angle=-90]{xmm_pps_radius.ps}
%\hspace{-4cm}
\end{center}
\caption{Evolution of the blackbody temperatures ({\em left}) and
  radii ({\em right}) as a function of the rotational phase for the {\XMM} observations. 
  Spectra of consecutive observations were coadded (obs. ID 0743630201-301 and 
  0743630401-501; see the two lower panels).}
  %Vertical dashed lines define the phase intervals used for the PPS analysis.
\label{fig:parametersvsphase}
\vskip -0.1truecm
\end{figure*}

The higher statistics of the {\XMM}/EPIC-pn data allowed us to put
more stringent constraints on the variations of the X-ray spectral
parameters along the spin phase. 
%For each of the three phase
%intervals, we fitted the spectra of all the observations jointly in the
%2--12 keV energy band, adopting the same prescriptions used for the
%phase-averaged spectroscopy in the filtering of the spectral
%channels. We applied only a BB+PL model, since the NTZ model is
%intrinsically phase-averaged. We tied the hydrogen
%column density and the power law photon indices to the best-fit values
%determined with the phase-averaged spectroscopy (see Table
%\ref{tab:XMMspectralfits}). We obtained $\chi^2_\nu =0.94$ for 182
%d.o.f. for the peak, $\chi^2_\nu =1.06$ for 103 d.o.f. for the
%secondary peak, and $\chi^2_\nu = 1.08$ for 102 d.o.f. for the pulse
%minimum. Spectral parameters for the peak and the minimum are reported
%in Table \ref{tab:pnspectrapps}. 
We extracted the background-subtracted spectra in six different 
phase intervals for each observation, as shown in Fig. \ref{fig:parametersvsphase}. We fitted all spectra with a BB+PL 
model, adopting the same prescriptions used for the phase-averaged 
spectroscopy in the filtering of the spectral channels. We tied the 
hydrogen column density and the power law photon indices to the 
best-fit values determined with the phase-averaged analysis
(see Table \ref{tab:XMMspectralfits}). We obtained statistically acceptable
results in all cases. The evolutions of the blackbody
temperature and emitting radius as a function of the rotational phase
for all the observations are shown in Fig.
\ref{fig:parametersvsphase}. Variability of both the parameters along
the rotational phase is more significant during the first observation
(a fit with a constant yields $\chi^2_\nu =2.6$ for 5 d.o.f. in both
cases) than in the following observations ($\chi^2_\nu \le 1.4$
for 5 d.o.f. in all cases).

\begin{figure*}
\begin{center}
\includegraphics[width=9cm]{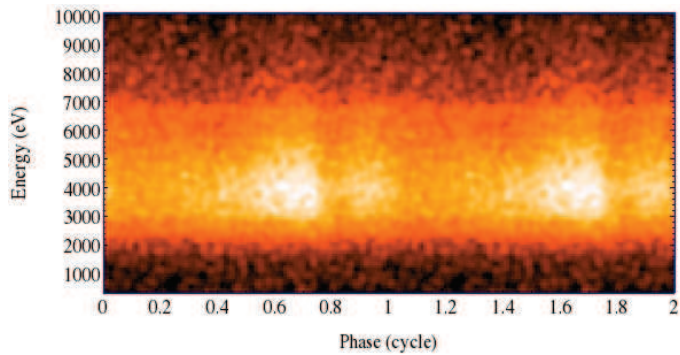}
%\hspace{-4cm}
\end{center}
\caption{Energy versus phase image for the {\XMM} 
  observation with the highest number of counts (obs. ID
  0724210201). The image was obtained by binning the EPIC-pn source 
  counts into 100 phase bins and energy channels of 100 eV, to better 
  visualize the shape of the pulse profile and its dependence on energy.}
\label{fig:evsphase}
\vskip -0.1truecm
\end{figure*}

To search for possible phase-dependent absorption features in the
X-ray spectra of \galcen\ (similarly to the one detected in \lowba;
Tiengo et al. 2013), we produced images of energy versus phase for each
of the eight EPIC-pn observations. We investigated different energy
and phase binnings. In Fig. \ref{fig:evsphase} we show 
the image for the observation with the highest number of counts 
(obs. ID 0724210201), produced by binning the source counts into 100 
phase bins and 100-eV wide energy channels. The spin period modulation 
is clearly visible, as well as the large photoelectric absorption below 2\,keV.
For all observations we then divided these values first by the average number of counts 
in the same energy bin and then by the corresponding 0.3--10 keV count rate in the same 
phase interval. No prominent features can be seen in any of the images.

%%%%%%%%%%%%%%%%%%%%%%%%%%%%%%%%%%%%%%%%%%%%%%%%%%%%%%%%%%%%%%%%%%%%%%%%%%%%%%%%%%%%%%%%%%
%%%%%%%%%%%%%%%%%%%%%%%%%%%%%%%%%%%%%%%%%%%%%%%%%%%%%%%%%%%%%%%%%%%%%%%%%%%%%%%%%%%%%%%%%

%%%%%%%%%%%%%%%%%%%%%%%%%%%%%%%%%%%%%%%%%%%%%%%%%%%%%%%%%%%%%%%%%%%%%%%%%%%%%%%%%%%%%%%%%%
%%%%%%%%%%%%%%%%%%%%%%%%%%%%%%%%%%%%%%%%%%%%%%%%%%%%%%%%%%%%%%%%%%%%%%%%%%%%%%%%%%%%%%%%%

\subsection{X-ray brightness radial profiles}

\begin{figure}
\begin{center}
\includegraphics[width=5.5cm,angle=-90]{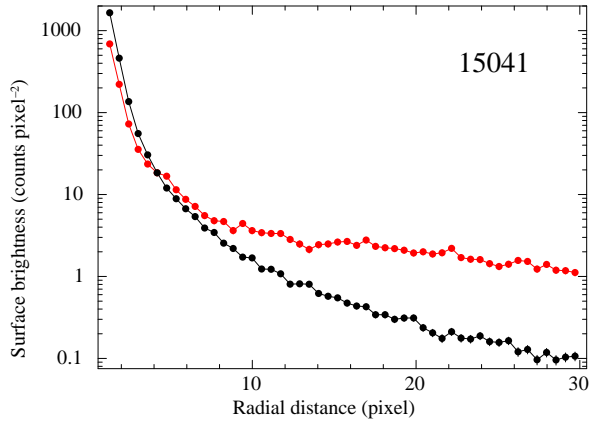}
\caption{Radial profile of the surface brightness for both the ACIS-S image of \galcen\, (red dots) and 
the \textsc{chart}/\textsc{marx} PSF plus a constant background (black dots) for the observation with the 
highest number of counts (obs. ID 15041). The simulated surface brightness has been normalized to 
match the observed one at 4 pixels (one ACIS-S pixel corresponds to 0.492 arcsec). Extended emission 
around  \galcen\ is clearly detected in all the observations.}
\label{fig:psf_profile}
%\label{fig:SB}
\end{center}
\end{figure}

For all the {\AXAF} observations, we used the Chandra Ray Tracer
(\textsc{chart}\footnote{http://cxc.cfa.harvard.edu/chart.}; Carter et
al. 2003) to simulate the best available PSF for \galcen, setting the
exposure time of each simulation equal to the exposure time of the
corresponding observation.  For the input spectrum in \textsc{chart}
we employed the blackbody spectrum of Table \ref{tab:spectralfits},
accounting for the pile-up. We then projected the PSF rays on to the
detector plane via the Model of AXAF Response to X-rays software
(\textsc{marx}\footnote{http://space.mit.edu/CXC/MARX.}, version
4.5.0; Wise et al. 2003). We extracted the counts of both the
simulated PSFs and the ACIS event files through 50 concentric annular
regions centred on the source position and extending from 1 to 30
pixels (1 ACIS-S pixel corresponds to 0.492 arcsec).
We then generated the X-ray brightness radial profiles and normalized the 
nominal one (plus a constant background) to match the observed 
one at a radial distance of 4 pixels, i.e. at a distance at which pile-up
effects are negligible. A plot of the observed and simulated 
surface brightness fluxes (in units of counts $\times$ pixel$^{-2}$) 
versus radial distance from the position of \galcen\ is shown in Fig. 
\ref{fig:psf_profile} for the observation with the highest number of counts 
(obs. ID 15041). 

Extended emission around \galcen\ is clearly 
detected in all the observations, and it is likely dominated by the intense 
Galactic Centre diffuse emission.
%For the longest observation we also built the surface brightness distributions 
%in the 0.3--4 and 4--8 keV energy intervals. We found comparable excess
%emissions in both cases, that excludes a scattering halo origin of this extended emission. 
%thus revealing a weak dependence of the extended emission on the photon energy.
A detailed analysis of the diffuse emission, including its spatial extension 
and spectral properties, is beyond the scope of this paper, and will be 
published in a subsequent work.
%Possible contributions may arise from the presence of a magnetically 
%powered nebula produced by outflows of relativistic charged particles from the magnetar, 
%or the scattering of X-rays by interstellar dust grains along the line of sight (producing a faint 
%and diffuse X-ray halo around the source). 
%In the latter scenario, the scattered radiation strongly depends on the photon energy: the flux is 
%predicted to scale with the photon energy as $F \propto E^{-2}$. Strong haloes 
%are thus expected to be more clearly detectable at soft X-ray energies.

\section{Discussion}

\subsection{Outburst evolution and comparison with other magnetars}

\begin{figure}
\begin{center}
\includegraphics[width=6cm,angle=-90]{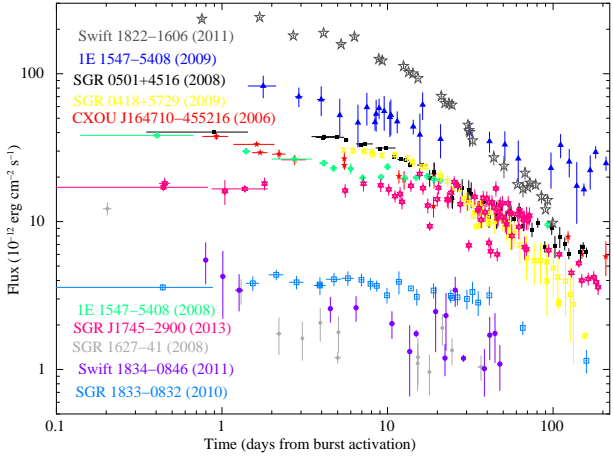}
\caption{Flux decay of all magnetar outbursts monitored with imaging instruments. Fluxes are absorbed in the 1--10 keV energy range (adapted and updated from Rea \& Esposito 2011).}
\label{fig:outbursts}
\end{center}
\end{figure}

The past decade has seen a great success in detecting magnetar outbursts, 
mainly thanks to the prompt response and monitoring of the {\em Swift} mission, 
and to the dedicated follow-up programs of \AXAF, \XMM , and more recently,
{\em NuSTAR}. The detailed study of about ten outbursts has shown many
common characteristics (see Rea \& Esposito 2011 for a review; see also
Fig.\,\ref{fig:outbursts}), although the precise triggering mechanism of these 
outbursts, as well as the energy reservoir responsible for sustaining the emission
over many months, remains uncertain.

All the outbursts that have been monitored with sufficient detail are
compatible with a rapid ($<$days) increase in luminosity up to a
maximum of a few $10^{35}$~erg~s$^{-1}$ and a thermally dominated
X-ray spectrum which softens during the decay. In the case of \sgre\ and
\aa, a non-thermal component extending up to 100--200 keV appears at
the beginning of the outburst, and becomes undetectable after
weeks/months (Rea et al. 2009; Bernardini et al. 2011; Kuiper et al. 2012).

The initial behavior of the 2013 outburst decay of \galcen\, was compatible 
with those observed in other magnetars. The outburst peak, the thermal 
emission peaked at about 1 keV, the small radiating surface (about 2\,km in 
radius), and the overall evolution in the first few months, were consistent with 
the behavior observed in other outbursts. However, after an additional year of X-ray 
monitoring, it became clear that the subsequent evolution of \galcen\, showed 
distinct characteristics. The source flux decay appears extremely slow: it is 
the first time that we observe a magnetar with a quiescent luminosity $<10^{34}$ erg 
s$^{-1}$ remaining at a luminosity $>10^{35}$ erg s$^{-1}$ for more than one 
year, and with a temperature decreasing by less than 10\% from the initial $\sim1$\,keV . 
A further interesting feature of this source is that the non-thermal component (as detected by {\XMM}) persisted on 
a very long temporal baseline during the outburst evolution. The flux due to the power law component does
not change significantly in time and, as a result, its fractional contribution to the total flux is larger at late times: 
$\sim520$ d after the outburst onset, $\lesssim50$ per cent of the 1--10 keV absorbed flux is due to the 
non-thermal component.

We first modelled the decay empirically to gauge the characteristic decay timescales. 
We adopted three different functions to model the blackbody temperature, 
radius and 1--10 keV absorbed flux temporal evolutions (see Fig. \ref{fig:parameters}): 
(i) a linear model; (ii) a power law: $f(t)=f_{0,PL}\times t^{-\Gamma}$; (iii) an
exponential: $f(t)=f_{0,exp}\times \rm{exp}[-(t-t_0)/\tau]$, where $t_0$ is the 
epoch of the first burst detected (which we fixed to 2013 April 24 in all cases) 
and $\tau$ is the e-folding time. 

The temporal evolution of the magnetar temperature is well represented by a linear 
model ($\chi^2_\nu =0.7$ for 23 d.o.f.), with initial temperature $kT_{BB,0}=0.85(1)$ 
keV and slope ($-1.77\pm0.04$) $\times$ 10$^{-4}$. The hot spot shrinking is best 
modeled by an exponential ($\chi^2_\nu =0.8$ for 23 d.o.f.). 
Best-fit parameters are $\tau=640\pm62$ d and initial radius 
$R_{BB,0}=2.60\pm0.08$ km. The shape of the flux decline appears to change 
in time and in fact none of these models can accurately describe the 
magnetar flux overall decay.  The flux decay during the first 100\,days since the outburst 
onset is well modelled by a linear plus exponential model with $\tau=37\pm2$ d 
($\chi^2_\nu$ = 1.5 for 4 d.o.f.). After $\sim100$ days, the best fitting model turns out to be an 
exponential with $\tau=253\pm5$ d ($\chi^2_\nu$ = 1.4 for 15 d.o.f.).

\subsection{Crustal cooling modelling}

We applied the crustal cooling model (see e.g. Pons \& Rea 2012)
to the data collected during the 1.5-yr outburst of \galcen . Although
this model was successful in explaining several other magnetar outbursts
(Rea et al. 2012, 2013b), in this case we could not reproduce the very
slow cooling and high luminosity observed for this source. We ran
several models varying the total injected energy, the angular size,
and the depth of the region where the energy is released, but we could
not find any set of parameters that fit the data. 

In the framework of the starquake model, the maximum temperature reached 
in the region where the energy is released is limited by neutrino emission 
processes. This internal temperature determines the maximum surface temperature 
and therefore the luminosity at which the outburst peaks during the
first few days. For injected energies $>10^{43}$\,erg, there is no significant 
increase in the peak luminosity because the crustal temperature saturates 
(at about $3-5\times10^{9}$\,K) due to the efficient neutrino processes. After
reaching the maximum luminosity (between 1 hour and 1 day depending on
the depth and injection rate), the cooling curve tracks the thermal
relaxation of the crust. Independent of the initial injected energy
and surface temperature, the luminosity is expected to drop below
$10^{35}$ erg s$^{-1}$ after $<$20-30 days (see e.g. Fig.\,1 in Pons
\& Rea 2012), due to neutrino emission processes in the crust (mainly
plasmon decay, and probably neutrino synchrotron for magnetar field
strengths).

In Fig.\,\ref{fig:cooling} (left panel, lower curves) we show an example of the
expected cooling curve of a magnetar with the same characteristics of
\galcen\,.  We assume that a sudden large energy release, $E
\simeq 10^{45}$ erg, heats up a layer of the outer crust up to
$3\times 10^9$\,K.  We also assume that the event affects the
entire magnetar surface, to create the most favorable scenario (the
luminosity simply scales with the area of the emitting region), and
that the layer where the energy is injected extends from an external
boundary at $\rho_{\rm OUT} \sim 3\times10^{9}$ g\,cm$^{-3}$, to an
inner boundary at $\rho_{\rm IN} \sim 2$ and $4\times10^{10}$ g\,cm$^{-3}$
(we show these two cases in the two lower curves of Fig.\,\ref{fig:cooling}, left panel). 
It is clear that, even in this most favorable
case, the high luminosities observed at late times are difficult to
reconcile with any cooling model. In particular, injecting more energy
or changing $\rho_{\rm OUT}$ will only affect the peak
luminosity during the first days or weeks. On the other hand, injecting
energy deeper into the crust (i.e. at higher $\rho_{\rm IN}$), is
expected to change the late time evolution only slightly. This can be
seen by comparing the solid and dashed lines in the left panel of
Fig.~\ref{fig:cooling}, which correspond to $\rho_{\rm IN}= 2$ and 4
$\times10^{10}$ g\,cm$^{-3}$, respectively.

%%%%%%%%%%%%%%%%%%%%%%%%%%%%%%%%%%%%%%%%%%%%%%%%%%%%%%%%%%%%%%%%%%%%%%%%%%%%%%%%%%%%%%%%%%
%%%%%%%%%%%%%%%%%%%%%%%%%%%%%%%%%%%%%%%%%%%%%%%%%%%%%%%%%%%%%%%%%%%%%%%%%%%%%%%%%%%%%%%%%%

For illustrative purposes, we also show the cooling curves obtained
when plasmon and synchrotron neutrino processes are switched off (see
the upper curves in the left panel of Fig.\,\ref{fig:cooling}). These provide 
a much closer match to the data; however, there is no clear reason why these
neutrino processes should not operate in these conditions. This
example is only meant to highlight the relevance of understanding
neutrino processes in the crust, especially under the presence of
strong fields. Another possibility to fit the data is to tune the
energy injection, which must be maintained during the first $\sim
200$\,days, resulting in a higher luminosity at late times. If we
assume that only a region 5 km in radius is affected (this is closer 
to the $\sim2$\,km emitting region observed), we need a
continuous injection of at least $\sim10^{44}$ erg s$^{-1}$ (per
day) for about 200\,days, which results in a total energy of a few
$10^{46}$\,erg. While this energy budget may not be unrealistic,
a physical mechanism that can operate for such a long timescale
is not known. A possibility might be a continuous injection of energy  
to keep the surface at high temperatures for
so long, although in this latter case we should possibly expect more SGR-like bursts during the first hundreds days.

%%%%%%%%%%%%%%%%%%%%%%%%%%%%%%%%%%%%%%%%%%%%%%%%%%%%%%%%%%%%%%%%%%%%%%%%%%%%%%%%%%%%%%%%%%
%%%%%%%%%%%%%%%%%%%%%%%%%%%%%%%%%%%%%%%%%%%%%%%%%%%%%%%%%%%%%%%%%%%%%%%%%%%%%%%%%%%%%%%%%%

\subsection{Bombardment by magnetospheric currents in a bundle}

In this section we discuss the possibility that the prolonged high
luminosity of \galcen\, is in part due to external particle bombardment as a
consequence of the existence of a twisted magnetic field bundle. A
valid alternative model to the crustal cooling scenario invokes
the presence of magnetospheric currents flowing along a gradually
shrinking magnetic bundle, and heating the surface from
outside. According to Beloborodov (2007, 2013), this bundle can 
untwist on different timescales: i) in the equatorial regions of
the magnetosphere, where the magnetic field reaches a few stellar radii,
currents are dissipated after weeks or months, while ii) at higher
latitudes (close to the poles), a bundle may untwist more
slowly, possibly in one to ten years.  Here, particles can reach
Lorentz factors of a few tens (Beloborodov 2007). In this scenario, a
quasi steady-state outflow of electrons and positrons is maintained
thanks to magnetic pair production close to the surface. The
non-negligible electric voltage along the magnetic field lines and the
radiative force due to Compton scattering regulate the
streams of positrons and electrons along the field line.

%%%%%%%%%%%%%%%%%%%%%%%%%%%%%%%%%%%%%%%%%%%%%%%%%%%%%%%%%%%%%%%%%%%%%%%%%%%%%%%%%%%%%%%%%%
\begin{figure*}
\begin{center}
\includegraphics[width=8.8cm]{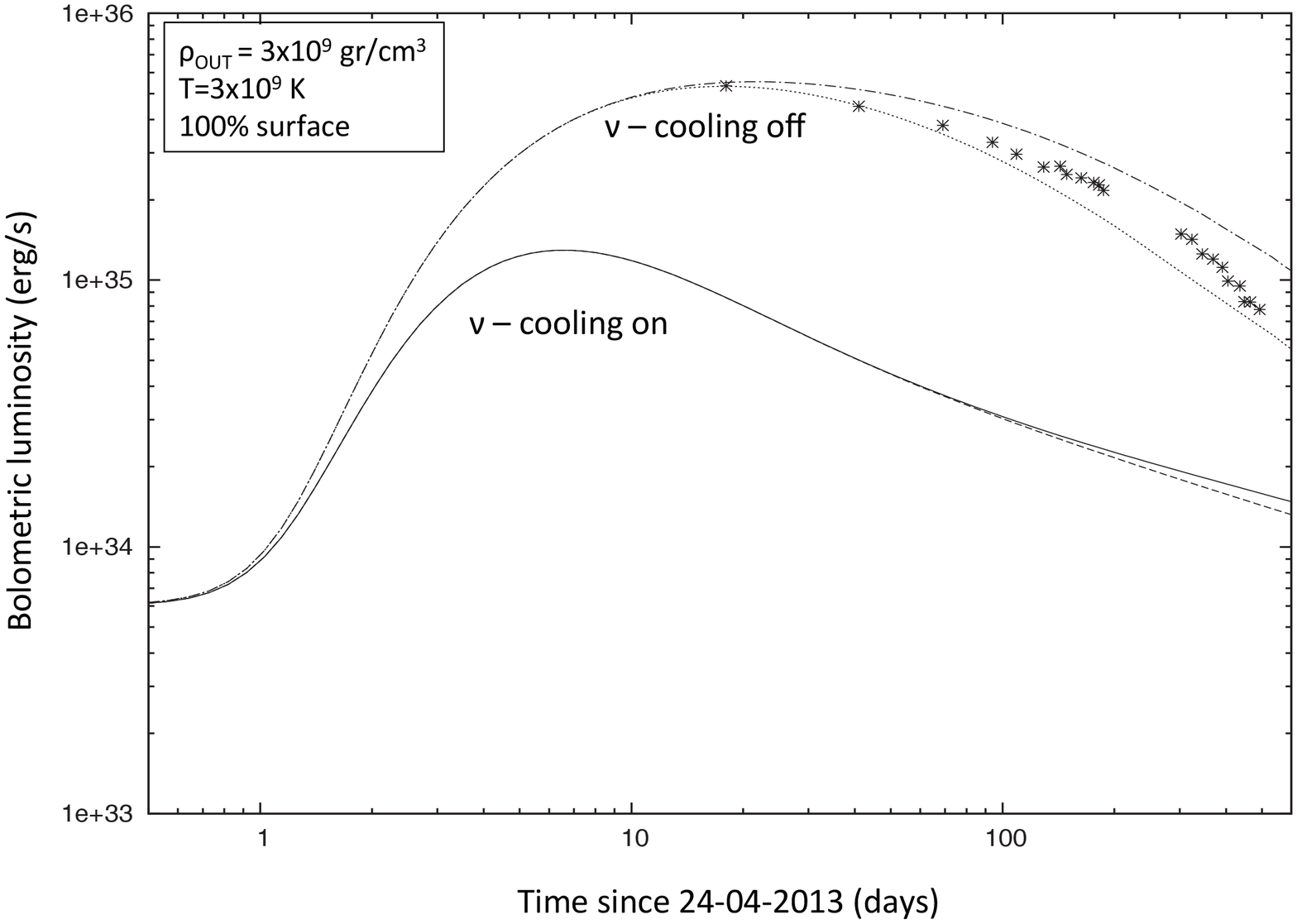}
\includegraphics[width=8.2cm]{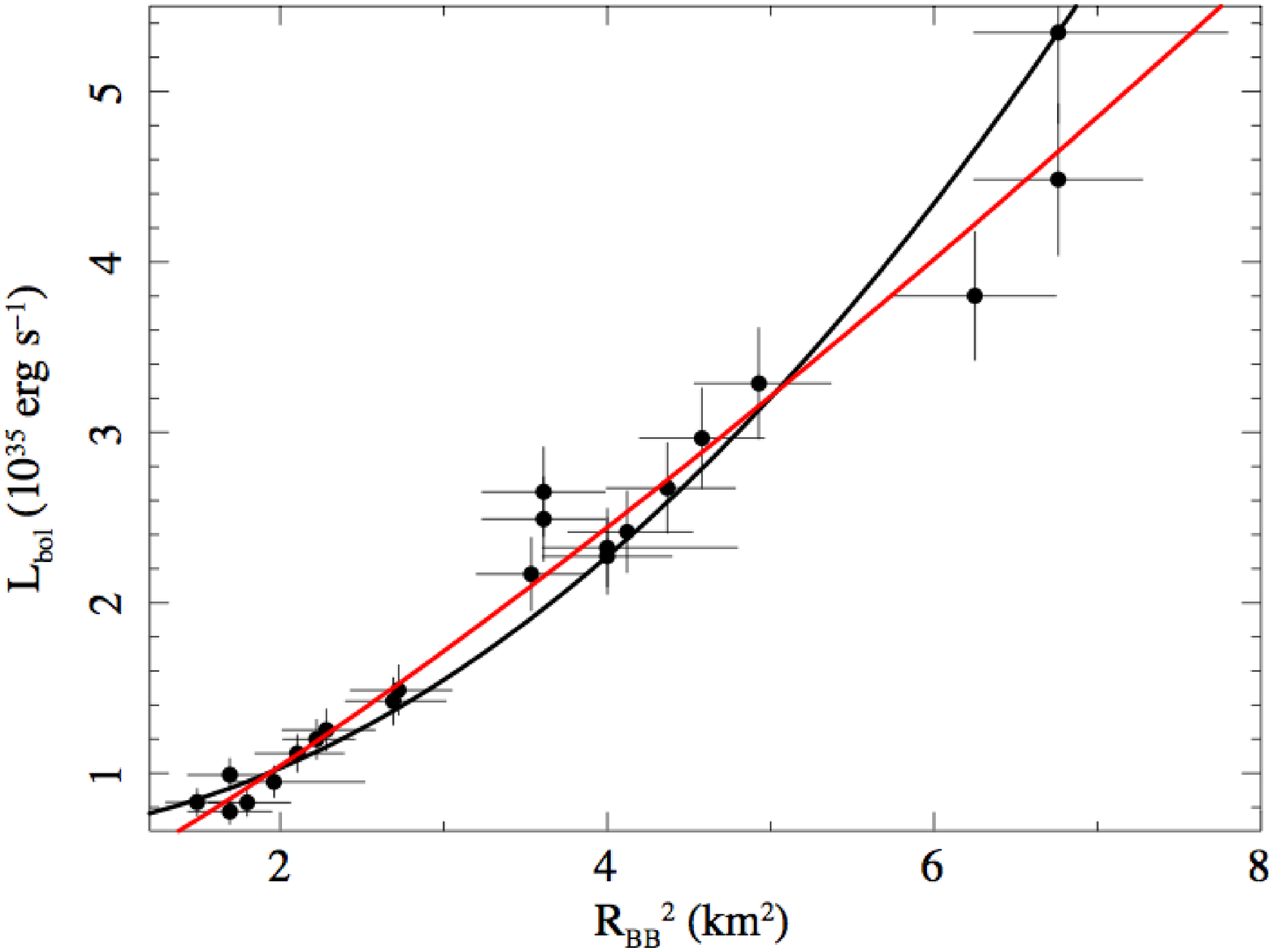}
\caption{{\em Left panel}: crustal
cooling curves attempting at modelling the luminosity decrease of \galcen . 
Luminosities are bolometric and calculated assuming a distance of
8.3\,kpc. For the neutrino-cooling on and off set of curves, the lower and upper curves are relative to $\rho_{\rm IN}=2$ and $4\times10^{10}$ g\,cm$^{-3}$, respectively. {\em Right panel}: bolometric luminosity as a
function of the square of the blackbody radius at infinity. Solid
lines represent the fits with a quadratic function (black) and a generic
power law ($\alpha=1.23(8)$; red).}
\label{fig:cooling}
\label{fig:lumradius}
\end{center}
\end{figure*}
%%%%%%%%%%%%%%%%%%%%%%%%%%%%%%%%%%%%%%%%%%%%%%%%%%%%%%%%%%%%%%%%%%%%%%%%%%%%%%%%%%%%%%%%%%

The presence of a non-thermal component observed by {\em NuSTAR} 
(Mori et al. 2013; Kaspi et al. 2014), and confirmed also by our \XMM\,
observations on a much longer temporal baseline, is suggestive
of a large density of magnetospheric particles which boost thermal photons 
emitted from the surface via resonant Compton scattering, providing the 
power law component. In this context, the observed $\sim$2\,km size of the 
emitting blackbody is consistent with a relatively small j-bundle. In the scenario 
in which the outburst evolution is dominated by an untwisting bundle and the 
poloidal magnetic field has a dipole geometry, the luminosity is expected 
to decrease with the square of the blackbody area ($A_b= 4 \pi R_{BB}^2$; 
Beloborodov 2007, 2009). A flatter dependence may arise from a more complex 
field geometry. In Fig.\,\ref{fig:lumradius} we show the fits of the bolometric 
luminosity as a function of $R_{BB}^2$ with two different models, a quadratic 
function $L_{\rm bol} \propto A_b^2$ (black line; $\chi_{\rm \nu}^2 = 1.3$ for 
23 d.o.f.) and a power law $L_{\rm bol} \propto A_b^{\alpha}$ (red line; 
$\chi_{\rm \nu}^2 = 0.8$ for 23 d.o.f.). For the latter model we find $\alpha=1.23(8)$. 
Interestingly, a similar relation was observed also for the outburst decay 
of \lowba\ (Rea et al. 2013b) and \wes\ (An et al. 2013).

In the following we will assess, using first order approximations, whether the particle density needed
to keep the footprint of the bundle at a temperature of $\sim$1\,keV
for the first hundreds of days after the outburst onset is consistent
with the particle density in the bundle responsible for the
non-thermal power law tail. The power of the infalling particles is
$E_{\rm kin} \dot{N}$, where $E_{\rm kin}$ is the kinetic energy of a
single particle at the surface and $\dot{N}$ is the total number of
infalling particles per unit time. If this kinetic energy is
transferred by the infalling particles to the footprint of the
bundle, and produces thermal luminosity from the footprint surface,
then:

\begin{equation}
  L_X = A_b \sigma T^4 = E_{\rm kin}\dot{N} = n\Gamma m_e c^3 A_b, 
\end{equation}
where $A_b$ is the area of the footprint surface, $T$ is the spot
temperature, $n$ is the density of the infalling particles (assumed
to be electrons and/or positrons, created by means of pair
production), and $\Gamma$ is the Lorentz factor. We calculated the
density of the infalling particles by considering the kinetic energy
they need to heat the base of the bundle spot. For a given temperature, 
one can estimate $n$ as

\begin{equation}\label{eq:n_bomb}
  n_{\rm bomb} = \frac{\sigma T^4}{m_e \Gamma c^3} \sim 4.2\times 10^{22}~\frac{[kT/(1\ {\rm keV})]^4}{\Gamma}  {\rm cm^{-3}}~.
\end{equation}

On the other hand, we can estimate the density of the particles responsible 
for the resonant Compton scattering which produces the X-ray tail as

\begin{equation}\label{eq:n_rcs}
  n_{\rm rcs} \simeq \frac{J_B {\cal M}}{v e} \simeq \frac{ {\cal M} 
B}{4\pi \beta e r} \sim 1.7 \times 10^{16} \frac{{\cal M} B_{14}}{ 
\beta } \left(\frac{r}{R_*}\right)^{-1}{\rm cm^{-3}}~, 
\end{equation} 
where 
$\vec{J_B}=(c/4\pi)\vec{\nabla}\times\vec{B}$ is the conduction current, 
$B$ is the local magnetic field, and $r$ is the length-scale over which 
$B$ varies ($R_*\sim 10^6 \ \rm {cm}$ is the star radius). 
In the magnetosphere of a magnetar the real current is always very close 
to $J_B$ and it is mostly conducted by e$^\pm$ pairs (Beloborodov 2007). 
The abundance of pairs is accounted for by the multiplicity factor ${\cal M}$ 
which is the ratio between the actual charge density (including pairs) and the 
minimum density needed to sustain $J_B$; the latter corresponds to a 
charge--separated flow in which the current is carried only by electrons (and 
ions). If the same charge population is responsible for both resonant Compton 
scattering and surface heating, the densities given by eqs.~(\ref{eq:n_bomb}) 
and (\ref{eq:n_rcs}) should be equal. This implies

\begin{equation}\label{eq:infer_b}
  B_{14}\left(\frac{r}{R_*}\right)^{-1} {\cal M} \Gamma = 2.5 \times 10^6 \left(\frac{kT}{1\ \rm {keV}}\right)^4\,.
\end{equation}

According to Beloborodov (2013), both the Lorentz factor and the pair
multiplicity change along the magnetic field lines, with typical values of
${\cal M}\sim 100$ (i.e. efficient pair creation), $\Gamma \sim 10$ in
the largest magnetic field loops, and ${\cal M} \sim 1$ (i.e.
charge-separated plasma), $\Gamma \sim 1$ in the inner part of the
magnetosphere. 
The previous equality cannot be satisfied for a typical
temperature of $\sim 0.8-1$ keV, unless the magnetic field changes
over an exceedingly small length-scale, a few meters at most. It
appears, therefore, very unlikely that a single flow can explain both
surface heating and resonant up-scattering.

\section{Conclusions}

The spectacular angular resolution of \AXAF\, and the large effective area of 
\XMM , together with an intense monitoring of the Galactic Centre region, has allowed 
us to collect an unprecedented dataset covering the outburst of \galcen , with very little 
background contamination (which can be very severe in this region of the Milky Way).

The analysis of the evolution of the spin period allowed us to find three different 
timing solutions between 2013 April 29 and 2014 August 30, which show that 
the source period derivative has changed at least twice, from 6.6$\times10^{-12}$ 
s~s$^{-1}$ in 2013 April at the outburst onset, to 3.3$\times10^{-11}$ s~s$^{-1}$ 
in 2014 August. While the first $\dot{P}$ change could be related with the occurrence 
of an SGR-like burst (Kaspi et al. 2014), no burst has been detected from the source 
close in time to the second $\dot{P}$ variation (although we cannot exclude it was 
missed by current instruments). This further change in the rotational 
evolution of the source might be related with the timing anomaly observed in the radio 
band around the end of 2013 (Lynch et al. 2015), unfortunately during our observing gap.

The 0.3--8\,keV source spectrum is perfectly modelled by a single blackbody with 
temperature cooling from $\sim$0.9 to 0.75\,keV in about 1.5 years. A faint non-thermal 
component is observed with \XMM . It dominates the flux at energies $>$8\,keV at all the 
stages of the outburst decay, with a power law photon index ranging from $\sim1.7$ to $\sim2.6$. 
It is most probably due to resonant Compton scattering onto non-relativistic electrons in the 
magnetosphere.

Modelling the outburst evolution with crustal cooling models has difficulty in explaining 
the high luminosity of this outburst and its extremely slow flux decay. If the outburst evolution 
is indeed due to crustal cooling, then magnetic energy injection needs to be continuous over 
at least the first $\sim$200 days. 

The presence of a small twisted bundle sustaining currents bombarding the surface region 
at the base of the bundle, and keeping the outburst luminosity so high, appears a viable scenario 
to explain this particular outburst. However, detailed numerical simulations are needed to 
confirm this possibility.

This source is rather unique, given its proximity to Sgr~A$^*$. In particular, it has a $>90$ per 
cent probability of being in a bound orbit around Sgr~A$^*$ according to our previous N-body 
simulations (Rea et al. 2013a), and the recent estimates inferred from its proper motion (Bower 
et al. 2015). We will continue monitoring the source with \AXAF\, and \XMM\, for the coming year.

\section*{Acknowledgements}
FCZ and NR are supported by an NWO Vidi Grant (PI: Rea) and by the European COST Action 
MP1304 (NewCOMPSTAR). NR, AP, DV and DFT acknowledge support by grants 
AYA2012-39303 and SGR2014-1073. AP is supported by a Juan de la Cierva fellowship. 
JAP acknowledges support by grant AYA 2013-42184-P.
PE acknowledges a Fulbright Research Scholar grant administered by the U.S.--Italy 
Fulbright Commission and is grateful to the Harvard--Smithsonian Center for Astrophysics
for hosting him during his Fulbright exchange. DH acknowledges support from {\em Chandra 
X-ray Observatory} (CXO) Award Number GO3-14121X, operated by the Smithsonian Astrophysical 
Observatory for and on behalf of NASA under contract NAS8-03060, and also by NASA \swift\ 
grant NNX14AC30G. GP acknowledges support via an EU Marie Curie Intra-European fellowship 
under contract no. FP-PEOPLE-2012-IEF- 331095 and the Bundesministerium f\"{u}r Wirtschaft und 
Technologie/Deutsches Zentrum f\"{u}r Luft-und Raumfahrt (BMWI/DLR, FKZ 50 OR 1408) and the 
Max Planck Society. RP acknowledges partial support by {\AXAF} grants (awarded by SAO) G03-13068A 
and G04-15068X. RPM acknowledges funding from the European Commission Seventh Framework 
Programme (FP7/2007-2013) under grant agreement n. 267251. FCZ acknowledges CSIC-IEEC for
very kind hospitality during part of the work and Geoffrey Bower for helpful discussions. 
The scientific results reported in this article are based on observations obtained with the 
{\em Chandra X-ray Observatory} and {\XMM}, an ESA science mission with instruments and 
contributions directly funded by ESA Member States and NASA. This research has made use 
of software provided by the {\AXAF} X-ray Center (CXC) in the application package CIAO, and 
of softwares and tools provided by the High Energy Astrophysics Science Archive Research 
Center (HEASARC), which is a service of the Astrophysics Science Division at NASA/GSFC and 
the High Energy Astrophysics Division of the Smithsonian Astrophysical Observatory.

%\label{lastpage}

\end{document}